\RequirePackage{fix-cm}
\documentclass[aap,preprint]{imsart}
\RequirePackage{amsthm,amsmath,amsfonts,amssymb,mathtools}
\RequirePackage[T1]{fontenc}
\RequirePackage{microtype,needspace,graphicx,xcolor}
\RequirePackage[numbers,sort&compress]{natbib}
\RequirePackage[colorlinks=true,linkcolor=blue!55!black,citecolor=blue!55!black,urlcolor=blue!55!black]{hyperref}
\DeclareFontShape{T1}{ptm}{m}{scit}{<->ssub * ptm/m/sc}{}
\makeatletter
\appto\ps@copyright{\let\@oddhead\@empty\let\@evenhead\@empty}
\makeatother
\hypersetup{pdftitle={Covariance spectra of random linear systems near instability},pdfauthor={Pengbin Feng},pdfsubject={Random matrices and critical stationary covariance},pdfkeywords={Independent entries, covariance spectrum, critical damping, circular element}}
\startlocaldefs
\numberwithin{equation}{section}
\theoremstyle{plain}
\newtheorem{theorem}{Theorem}[section]
\newtheorem{lemma}[theorem]{Lemma}
\newtheorem{proposition}[theorem]{Proposition}
\newtheorem{corollary}[theorem]{Corollary}
\theoremstyle{definition}

\newcommand{\R}{\mathbb R}
\newcommand{\N}{\mathbb N}
\newcommand{\E}{\mathbb E}
\newcommand{\Prob}{\mathbb P}
\newcommand{\Law}{\operatorname{Law}}
\newcommand{\Beta}{\operatorname{Beta}}
\newcommand{\tr}{\operatorname{Tr}}
\newcommand{\spec}{\operatorname{spec}}
\newcommand{\one}{\mathbf 1}
\newcommand{\dd}{\,\mathrm d}
\newcommand{\norm}[1]{\lVert #1\rVert}
\newcommand{\abs}[1]{\lvert #1\rvert}
\newcommand{\Cat}{\operatorname{Cat}}

\endlocaldefs
\begin{document}
\begin{frontmatter}
\title{Covariance spectra of random\\linear systems near instability}
\pdftitle{Covariance spectra of random linear systems near instability}
\pdfsubject{Random matrices and stationary covariance near instability}
\runtitle{Covariance spectra near instability}
\begin{aug}
\author[A]{\fnms{Pengbin}~\snm{Feng}\ead[label=e1]{pengbinf@alumni.usc.edu}}
\address[A]{University of Southern California\printead[presep={,\ }]{e1}}
\end{aug}

\begin{abstract}
We study how stationary variance is distributed among the principal
components of a randomly coupled linear system near instability.
The coupling matrix has entries $N^{-1/2}\xi_{ij}$, where the
$\xi_{ij}$ are independent copies of a bounded, centered real random
variable of variance one. As $N\to\infty$ and the excess damping
$\delta_N\to0$, we prove that the covariance spectrum, with eigenvalues
rescaled by $\delta_N^2$ and weighted by their shares of the total
variance, converges in probability to an explicit beta-product law.
The convergence holds throughout $\delta_N\ge N^{-1/2+\varepsilon}$
for every fixed $0<\varepsilon<1/2$. The law describes eigenvalues
of order $\delta_N^{-2}$ at ranks of order $N\delta_N^{3/2}$.
Its limiting rank profile decays as $r^{-4/3}$ at large normalized
rank and determines how many principal components retain a prescribed
fraction of the stationary variance. The proof combines an exact
computation of planar moments, uniform estimates for growing matrix
words that accommodate nonsymmetric entry distributions, and resolvent
bounds controlling the contribution of long times.
\end{abstract}

\begin{keyword}[class=MSC]
\kwdgroup[type=primary]{\kwd{60B20}}
\kwdgroup[type=secondary]{\kwd{46L54}}
\end{keyword}
\begin{keyword}
\kwd{Random matrices}
\kwd{stationary covariance}
\kwd{critical damping}
\kwd{circular element}
\end{keyword}
\end{frontmatter}

\section{Introduction}

Small disturbances are transmitted repeatedly through a network
before they decay. Near a stability threshold, their accumulated
response can be large and concentrated along a small fraction of the
directions. The covariance spectrum records this concentration:
it measures the variance in each principal direction and determines
how many directions are needed to represent the fluctuations.
We study this spectrum for the linear stochastic system
\begin{equation}\label{eq:OU}
 dY_t=\bigl(X_N-(1+\delta)I\bigr)Y_t\dd t+dB_t,
\end{equation}
where $B_t$ is standard $N$-dimensional Brownian motion, independent
of $X_N$. When the drift is stable, its stationary covariance,
conditional on $X_N$, solves
\[
 2(1+\delta)Q-X_NQ-QX_N^{\mathsf T}=I.
\]
The trace of $Q$ is the total stationary noise energy. Its eigenvalues
are the variances of the principal components. If $X_N$ is nonnormal,
nonorthogonal modes can reinforce one another over finite times even
though every mode eventually decays. The resulting covariance can
therefore have a broad spectrum whose scale is not apparent from
the eigenvalues of $X_N$ alone.

This question has a direct interpretation in the analysis of neural
population activity, where principal component spectra describe how
variance is shared among collective activity patterns
\cite{HS,PZG26,ZV}. Given a high-dimensional stationary state, one
may wish to retain a prescribed fraction of its variance using as few
principal components as possible. A covariance spectral law determines that
number and the mean-square error of the corresponding projection.
For random recurrent dynamics, it also gives a baseline for the
concentration of activity produced by repeated interactions near
instability.

Financial networks provide a related motivation.
Dynamic contagion models for systemic risk are developed in
\cite{feng2021dynamic}. In \cite{feng2026lowrankgraphonlimitsdynamic},
finite-factor exposure matrices reduce threshold contagion to a finite
feedback system, while directed kernels describe its continuum limit.
These models describe how structured exposures generate collective
feedback. Here we ask how strongly repeated interactions can
concentrate fluctuations when the coupling entries are independent.

Our main result identifies the amplified part of the covariance
spectrum in a simultaneous large-matrix and critical-damping limit.
A single explicit profile describes its eigenvalues and their ranks
for every bounded, centered real iid interaction law of variance one,
including skewed laws. The limiting rank curve has exponent $4/3$.
The same profile gives the number of principal components needed
to retain any fixed fraction of the stationary variance.

We take $X_N=N^{-1/2}(\xi_{ij})$, where all $N^2$ entries, including
the diagonal, are independent copies of a real random variable $\xi$.
Throughout the paper,
\begin{equation}\label{eq:entry-assumptions}
 \E\xi=0,\qquad \E\xi^2=1,\qquad |\xi|\le M
 \quad\hbox{almost surely},
\end{equation}
where $M\ge1$ is fixed. The distribution does not depend on $N$.
Write $\tau_N=N^{-1}\tr$ and define
\begin{equation}\label{eq:Qdef}
 Q_{N,\delta}=
 \begin{cases}
 \displaystyle\int_0^\infty e^{-2(1+\delta)t}
                  e^{tX_N}e^{tX_N^{\mathsf T}}\dd t,
       &\max\Re\spec(X_N)<1+\delta,\\[2pt]
 0,&\text{otherwise}.
 \end{cases}
\end{equation}
The finite-time covariance $Q_{N,\delta,T}$ is the integral over
$[0,T]$, defined for every sample. In the regime considered below,
the stationary integral exists with probability tending to one.

To resolve the large covariance eigenvalues, introduce the finite
measure
\begin{equation}\label{eq:nuN}
 \nu_{N,\delta}(f)=\delta^{1/2}\tau_N
       [Q_{N,\delta}f(\delta^2Q_{N,\delta})].
\end{equation}
The factor $Q_{N,\delta}$ weights each direction by its contribution
to the stationary energy. The first-moment limit in
Theorem~\ref{thm:matrix-crossover} below gives
\[
 \tau_N(\delta_N^2Q_{N,\delta_N})
 =\delta_N^{3/2}\bigl(\delta_N^{1/2}\tau_NQ_{N,\delta_N}\bigr)
 \xrightarrow{\Prob}0
\]
under its assumptions. Thus the ordinary empirical measure on this
scale converges to the point mass at zero. Energy weighting retains
the distribution of the directions carrying the leading response;
the function $H$ below removes the weight to recover their counts.

\begin{theorem}\label{thm:matrix-crossover}
Assume \eqref{eq:entry-assumptions}. Let $\delta_N>0$ be deterministic
and tend to zero. Suppose that, for some fixed $0<\varepsilon<1/2$,
\begin{equation}\label{eq:window}
 \delta_N\ge N^{-1/2+\varepsilon}
 \quad\hbox{for all sufficiently large }N.
\end{equation}
Let $U\sim\Beta(1/4,3/4)$ and $V\sim\Beta(3/4,5/4)$ be independent,
and set
\[
 Y=UV/2,\qquad \nu_*=(2\sqrt2)^{-1}\Law(Y),\qquad
 H(x)=\int_{(x,\infty)}y^{-1}\nu_*(\dd y).
\]
Then $\nu_{N,\delta_N}\Rightarrow\nu_*$ in probability in the weak
topology of finite measures on $[0,\infty)$. In particular,
\begin{equation}\label{eq:first-mass}
 \delta_N^{1/2}\tau_NQ_{N,\delta_N}
       \xrightarrow{\Prob}(2\sqrt2)^{-1}.
\end{equation}
For every $0<a<b<\infty$,
\begin{equation}\label{eq:counts}
 \sup_{x\in[a,b]}\left|
 \frac{\#\{j:\lambda_j(Q_{N,\delta_N})>x\delta_N^{-2}\}}
      {N\delta_N^{3/2}}-H(x)\right|\xrightarrow{\Prob}0.
\end{equation}
List the eigenvalues in decreasing order. If deterministic integers
$k_N$ satisfy $k_N/(N\delta_N^{3/2})\to r\in(0,\infty)$, then
\begin{equation}\label{eq:quantile}
 \delta_N^2\lambda_{k_N}(Q_{N,\delta_N})
       \xrightarrow{\Prob}H^{-1}(r).
\end{equation}
Here $H:(0,1/2)\to(0,\infty)$ is continuous and strictly decreasing.
\end{theorem}

The spectral scale $\delta_N^{-2}$ and the rank scale
$N\delta_N^{3/2}$ together give a total stationary energy of order
$N\delta_N^{-1/2}$. The beta-product law specifies how this energy is shared
among the principal components.

The limiting rank profile has a power-law tail. Corollary~\ref{free:cor:ends}
gives
\begin{equation}\label{eq:rank-profile}
 H^{-1}(r)\sim\left(\frac{c_0}{r}\right)^{4/3},\qquad
 c_0=\frac{2^{9/4}}{9\sqrt\pi\,\Gamma(3/4)^2},
 \qquad r\to\infty.
\end{equation}
In \eqref{eq:quantile}, the matrix limit is taken at fixed normalized
rank $r$; \eqref{eq:rank-profile} describes the subsequent large-$r$
behavior of that curve.

The continuous-time law has an explicit relation to the discrete-time
critical profile computed in \cite{ZV}. After normalizing each energy
measure to unit mass and rescaling eigenvalues by the square of the
stability gap, Proposition~\ref{free:prop:discrete-profile} identifies
the continuous-time variable as the discrete-time variable multiplied
by the square of an independent arcsine variable with law
$\Beta(1/2,1/2)$. The discrete energy density
has an $x^{-1/2}$ singularity at the origin; the continuous density has
an $x^{-3/4}$ singularity. These yield rank exponents $2$ and $4/3$,
respectively.

\begin{corollary}\label{cor:explained-variance}
Under the assumptions of Theorem~\ref{thm:matrix-crossover}, set
$m_N=N\delta_N^{3/2}$. On stable samples, define
\[
 F_N(k)=\frac{\sum_{j=1}^{k}\lambda_j(Q_{N,\delta_N})}
                   {\tr Q_{N,\delta_N}},\qquad k\in\{0,\ldots,N\},
\]
with the empty sum equal to zero. For $0<\theta<1$, let
\[
 K_N(\theta)=\min\{k\in\{0,\ldots,N\}:F_N(k)\ge\theta\}.
\]
Set both quantities to zero on unstable samples. If deterministic
integers $k_N\in\{0,\ldots,N\}$ satisfy $k_N/m_N\to r\in(0,\infty)$,
then
\begin{equation}\label{eq:explained-variance}
 F_N(k_N)\xrightarrow{\Prob}\mathcal E(r)
       :=\Prob\{Y>H^{-1}(r)\}.
\end{equation}
The function $\mathcal E$ is continuous and strictly increasing,
with $\mathcal E(0+)=0$ and $\mathcal E(\infty)=1$. For each fixed
$0<\theta<1$,
\begin{equation}\label{eq:retained-dimension}
 \frac{K_N(\theta)}{m_N}\xrightarrow{\Prob}
 r_\theta:=\mathcal E^{-1}(\theta).
\end{equation}
\end{corollary}

Numerical evaluation of the beta-product law gives
$\mathcal E(1)\simeq0.2846$: the leading $N\delta_N^{3/2}$ principal
components retain about $28\%$ of the variance in the limit.
Retaining $90\%$ requires asymptotically
$r_{0.9}N\delta_N^{3/2}$ components, with
$r_{0.9}\simeq937.1$. The required number of components tends to
infinity, while its ratio to $N$ tends to zero. More generally,
$1-\mathcal E(r)$ is the limiting
optimal relative mean-square error of a rank-$k_N$ orthogonal
projection chosen for the given covariance. Every interaction law
satisfying \eqref{eq:entry-assumptions} has the same limiting
compression curve.

The remaining variance decays slowly along this curve. After the
matrix limit at each fixed $r$, the endpoint asymptotics of
Corollary~\ref{free:cor:ends} give
\begin{equation}\label{eq:compression-tail}
 1-\mathcal E(r)\sim6\sqrt2\,c_0^{4/3}r^{-1/3},
 \qquad r\to\infty,
\end{equation}
where $6\sqrt2\,c_0^{4/3}\simeq0.9831$. An eightfold increase in
normalized rank therefore asymptotically halves the unexplained
variance of the limiting profile.

\begin{figure}[t]
\centering
\includegraphics[width=\linewidth]{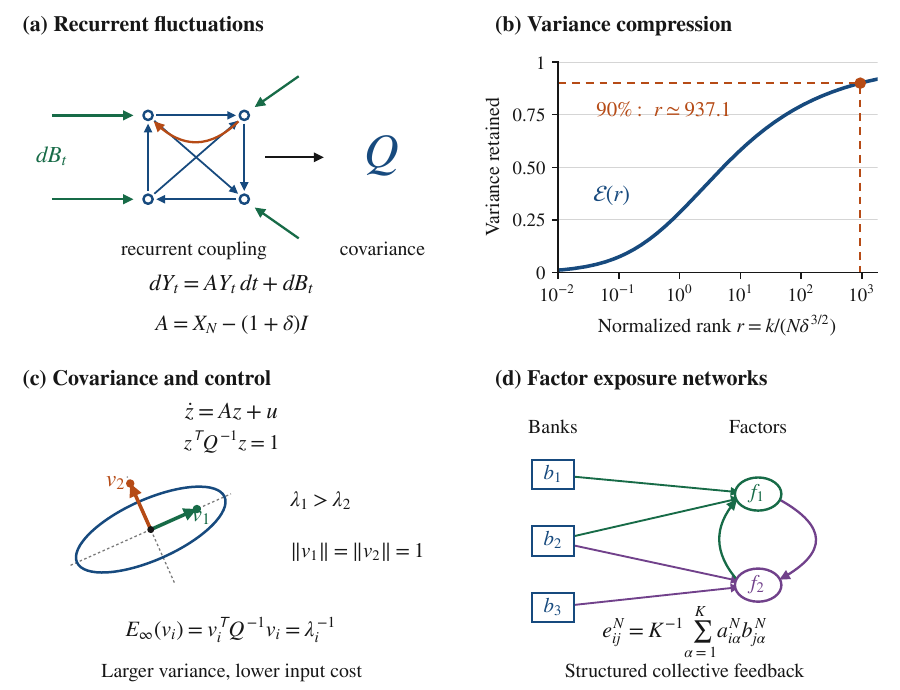}
\caption{Covariance spectra and collective response.
(a) Recurrent coupling transforms independent noise into fluctuations
along collective directions.
(b) The limiting variance fraction $\mathcal E(r)$ retained by the first
$k_N$ principal components when $k_N/(N\delta_N^{3/2})\to r$.
Numerical evaluation of the beta-product law gives $r_{0.9}\simeq937.1$.
(c) The covariance contour $z^{\mathsf T}Q^{-1}z=1$ and two unit
eigenvectors. With full-coordinate input, the infinite-horizon minimum
cost of reaching $v_i$ is $\lambda_i^{-1}$.
(d) Sender and receiver exposure factors in
\cite{feng2026lowrankgraphonlimitsdynamic} reduce threshold contagion
to a finite system of feedback coordinates.}
\label{fig:applications}
\end{figure}

The covariance also has a control interpretation. For each fixed stable
realization, consider the system with full-coordinate input,
$\dot z=(X_N-(1+\delta)I)z+u$. Its infinite-horizon controllability
Gramian is $Q_{N,\delta}$ \cite{PR16}.
Along a unit covariance eigenvector with variance $\lambda$, the
minimum input energy $\int_0^T\|u(t)\|^2\dd t$ needed to reach that
vector from the origin tends to $1/\lambda$ as $T\to\infty$.
The directions carrying large stationary fluctuations are therefore
the easiest to excite under this actuation.
Section~\ref{sec:transfer} gives the variational proof.
Figure~\ref{fig:applications} shows the recurrent model, its compression
curve and control interpretation, alongside the factor structure of
the financial-network background.

The circular model also permits a precise description of the upper
edge. If $c$ is a standard circular element and
\[
 q_\delta=\int_0^\infty e^{-2(1+\delta)t}e^{tc}e^{tc^*}\dd t,
\]
then Theorem~\ref{edge:thm:free-norm} proves
$\delta^2\norm{q_\delta}\to1/2$, the upper endpoint of the limiting
beta-product law.

\subsection{Previous work}

Erd\H{o}s, Kr\"uger and Renfrew \cite{EKR18,EKR23} analyze traces
of the form $\tau_N[f(X_N)g(X_N^*)]$ for independent entries and
elliptic correlations. Their results describe the critical decay of
the averaged response and the divergence of the first covariance
moment. The spectrum requires traces of products of response factors
at every fixed order. Fyodorov, Gudowska-Nowak, Nowak and Tarnowski
\cite{FGNT} study the equal-time Lyapunov covariance and its relation
to eigenvector nonorthogonality. Hu and Sompolinsky \cite{HS}
determine the spectrum of long-window covariance and discuss its
frequency dependence. For equal-time covariance, the matrix-valued
frequency response is integrated before its spectrum is taken.

Pachitariu et al.\ \cite{PZG26} use stationary covariance spectra
of randomly coupled linear systems to model principal component
spectra in neural recordings. They obtain a rank exponent $2/3$
for symmetric connections and report a numerical exponent near $1.25$
for nonsymmetric connections, leaving its mathematical derivation open.
Their simulations set the diagonal to zero and rescale each matrix
so that the largest real part of its eigenvalues is $0.998$;
our theorem follows a deterministic damping sequence in
\eqref{eq:window}. Preciado and Rahimian \cite{PR16} derive Gramian
spectral laws for symmetric random state matrices. Ferreira, Metz
and Barucca \cite{FMB25} study covariance spectra of reversible
Ornstein--Uhlenbeck systems with heterogeneous temperatures, whose
precision matrices form a deformed Wigner ensemble.

For the discrete-time quenched Gramian with real Gaussian interactions,
Zavatone-Veth \cite{ZV}
derives a scalar functional equation and a critical profile by a
formal free-probability calculation. It takes $N\to\infty$ at a fixed
variance parameter $g<1$, followed by $g\uparrow1$.
The coefficients of its critical
solution $1-\sqrt{1-2z}$ are the same Catalan integrals $J_p$ in
\eqref{free:eq:J}. In continuous time, individual response factors
also contain unequal powers of $c$ and $c^*$, and the corresponding
equations form an infinite hierarchy. We evaluate every fixed-order
critical moment by controlling these unequal blocks and their global
balance constraint. Both explicit laws can be expressed using the
classical beta-product family \cite{Dufresne}.

Our planar calculation uses the prescribed-block recursion of Kemp,
Mahlburg, Rattan and Smyth \cite{KMRS}. The organization of Gaussian
word expansions by surfaces is developed in \cite{DP,CF,EK}.
For bounded symmetric entries, we refine an index assignment by all
compatible Wick pairings and sum the cost of repeated vertex labels.
This gives uniform trace bounds at growing word lengths. Han
\cite[Theorem~2.3 and Appendix~C]{Han} treats nonsymmetric entry
distributions in high moments of the two-block word $X^k(X^*)^k$
on the scale $k=o(\sqrt N)$.
Earlier high-moment arguments treat nonsymmetric Wigner entry laws
\cite{PS07} and spectral-radius bounds for non-Hermitian matrices
with symmetric entry laws \cite{BCCT18}.
The covariance moments here involve arbitrary alternating block
lengths and connected products of traces. A pair--triple refinement gives
uniform counts for those prescribed blocks and for connected pairs of
traces. The exterior local laws of Alt, Erd\H{o}s and Kr\"uger \cite{AEK18,AEK21}
provide the stationary energy estimate.

\subsection{The argument}

The proof follows the dynamics over times of order $\delta_N^{-1}$,
when much of the stationary response accumulates. Expanding the matrix
exponentials produces words whose lengths grow with $N$.
Section~\ref{sec:free} computes the circular critical moments and
identifies the beta-product measure. Section~\ref{sec:comparison}
establishes the growing-time estimates for symmetric entries. Planar
trees with distinct vertex labels give the leading contribution;
collisions and connected pairings control the error and variance.
Section~\ref{sec:nonsymmetric} handles odd multiplicities through pair
and triple refinements and obtains the finite-time limits for general
bounded entries. The count retains the original block-length
constraints, giving the quadratic word-length scale needed for
\eqref{eq:window}.

Section~\ref{sec:real-energy} bounds the stationary energy by integrating
an exterior resolvent estimate. On one event of high probability,
the energy remaining after time $L/\delta_N$, normalized by
$\delta_N^{1/2}/N$, is at most $Ce^{-L}$.
Section~\ref{sec:transfer} takes the matrix limit at fixed
$L$ and then removes the time cutoff to obtain the spectral counts
and quantiles, followed by the variance fractions and projection
errors in Corollary~\ref{cor:explained-variance}.

Appendices~\ref{sec:gaussian} and \ref{app:physical-reconstruction}
prove the Gaussian and odd-pattern counts used in the comparison.
For the latter, a reversible compression of the minimal endpoint
quotient keeps the original block-length constraints inside the core
sum. Appendix~\ref{sec:free-edge} proves the circular norm limit.
A common Poisson translation bounds moments uniformly in their order,
and the semicircular Haagerup inequality \cite{Bozejko,Parraud}
turns that bound into an operator norm estimate.

\section{The circular covariance}
\label{sec:free}

Let $(\mathcal M,\tau)$ be a finite von Neumann algebra with faithful
normal tracial state, $\tau(1)=1$, and let $c\in\mathcal M$ be a standard
circular element, normalized by $\tau(cc^*)=1$. Define
\begin{equation}\label{free:eq:q}
 q_\delta=\int_0^\infty e^{-2(1+\delta)t}e^{tc}e^{tc^*}\dd t,
 \qquad \delta>0.
\end{equation}
The integral converges in norm by Lemma~\ref{free:lem:norm} and defines
a positive operator. Write $\mu_\delta$ for its spectral probability
measure with respect to $\tau$. Its mean diverges as
$\tau(q_\delta)\sim(2\sqrt{2\delta})^{-1}$, while the upper spectral
tail extends to the larger scale $\delta^{-2}$. Weighting the spectrum
by the eigenvalue retains the energy carried on this scale. We therefore
define the finite measure $\nu_\delta$ by
\begin{equation}\label{free:eq:nu}
 \int f(x)\,\nu_\delta(\dd x)
 =\delta^{1/2}\tau\bigl(q_\delta f(\delta^2q_\delta)\bigr)
\end{equation}
for bounded Borel functions $f$.

\begin{theorem}\label{free:thm:main}
For every fixed integer $p\geq1$,
\begin{equation}\label{free:eq:moments}
 \lim_{\delta\downarrow0}\delta^{2p-3/2}\tau(q_\delta^p)
 =C_p:=\frac{\Gamma(2p-\tfrac32)}
 {2^{3p-3/2}\sqrt\pi\,p!(p-1)!}.
\end{equation}
Let $U\sim\Beta(1/4,3/4)$ and $V\sim\Beta(3/4,5/4)$ be independent
classical random variables, and put $X=UV/2$. Then
\begin{equation}\label{free:eq:betalaw}
 \nu_\delta\ \Longrightarrow\ \nu_*:=\frac1{2\sqrt2}\Law(X).
\end{equation}
For every $x>0$,
\begin{equation}\label{free:eq:tail}
 \delta^{-3/2}\mu_\delta\bigl((x\delta^{-2},\infty)\bigr)
 \longrightarrow H(x):=\frac1{2\sqrt2}
             \E\bigl[X^{-1}\one_{\{X>x\}}\bigr].
\end{equation}
This convergence is locally uniform in $x\in(0,\infty)$. The function
$H$ is continuous, strictly positive on $(0,1/2)$, and zero on
$[1/2,\infty)$.
\end{theorem}

The law describes both the scale and the shape of the damping cutoff:
spectral values of order $\delta^{-2}$ carry a fraction of order
$\delta^{3/2}$ of the spectral measure. The first three moment constants are
\begin{equation}\label{free:eq:checks}
 C_1=\frac1{2\sqrt2},\qquad
 C_2=\frac3{128\sqrt2},\qquad
 C_3=\frac{35}{8192\sqrt2}.
\end{equation}

\begin{corollary}\label{free:cor:ends}
The crossover function has the asymptotics
\begin{align}
 H(x)&\sim\frac{2^{9/4}}{9\sqrt\pi\,\Gamma(3/4)^2}\,x^{-3/4}
       &&(x\downarrow0),\label{free:eq:small}\\
 H(x)&\sim\frac1\pi\left(\frac12-x\right)^2
       &&(x\uparrow1/2).\label{free:eq:edge}
\end{align}
Its inverse satisfies $H^{-1}(r)\sim(c_0/r)^{4/3}$ as $r\to\infty$,
where $c_0$ is the constant in \eqref{free:eq:small}.
\end{corollary}

The proof starts from the exact recursion for locally symmetric
words in \cite[Theorem~3.6]{KMRS}, with the cyclic order of the blocks
preserved. An insertion argument compares unequal blocks with these
symmetric words. Exponential averaging then reduces the leading
recursion to a scalar equation, from which all the required simplex
integrals can be evaluated. The time integrations become Poisson
averages: the total number of positive and negative letters must
agree, and the probability of this single constraint supplies the
square-root factor in the moment asymptotics. We take each limit at a
fixed moment order. The common compact support of the rescaled
measures then turns these moment limits into the spectral law.

\subsection{Planar pairings with specified block lengths}

For $a,b\in\N_0^p$, let $P_p(a,b)$ count the noncrossing pairings of
\begin{equation}\label{free:eq:word}
 +^{a_1}-^{b_1}\cdots+^{a_p}-^{b_p}
\end{equation}
in which each pair joins opposite signs. Set $P_p(a,b)=0$ when
$\sum a_i\ne\sum b_i$. The empty word has one pairing. Displayed
blocks of length zero are retained as marked intervals, so that their
positions remain specified even when the literal word has fewer runs.
Write
\[
 \Phi_p(\ell)=P_p(\ell,\ell),\qquad \ell\in\N_0^p.
\]

\subsubsection{The locally symmetric leading coefficient}

Let $\mathcal C_p$ be the set of decompositions of the ordered list
$(2,\ldots,p)$ into nonempty consecutive intervals
$I_1,\ldots,I_k$, where $1\leq k\leq p-1$. If the first coordinate
$\ell_1$ is minimal, the recursion of
\cite[Theorem~3.6]{KMRS} reads
\begin{equation}\label{free:eq:exact-rec}
 \Phi_p(\ell)=\sum_{(I_1,\ldots,I_k)\in\mathcal C_p}
       \binom{\ell_1+1}{k}\prod_{j=1}^k
                      \Phi_{|I_j|}(\ell_{I_j}),\qquad p\geq2,
\end{equation}
with $\Phi_1=1$. A cyclic rotation brings a minimum to the first
coordinate. Each recursive call makes its own cyclic rotation.
For a zero minimum, only the term $k=1$ remains, and
\eqref{free:eq:exact-rec} simply deletes the empty pair of blocks; thus
the formula also holds for nonnegative lengths.

Define $F_1=1$. For $p\geq2$ and $u\in\R_+^p$, rotate a minimum
to the first coordinate and set
\begin{equation}\label{free:eq:F-rec}
 F_p(u)=\sum_{(I_1,\ldots,I_k)\in\mathcal C_p}
          \frac{u_1^k}{k!}\prod_{j=1}^kF_{|I_j|}(u_{I_j}).
\end{equation}
Here and below $\R_+=[0,\infty)$. The next lemma shows that this
definition is independent of the choice among tied minima.

\begin{lemma}\label{free:lem:F}
The function $F_p$ is nonnegative, continuous, cyclically invariant,
and homogeneous of degree $p-1$. It is polynomial on each coordinate
ordering chamber. If $p\geq2$, it vanishes when any coordinate is zero,
and there is a constant $K_p<\infty$ such that
\begin{align}
 \abs{\Phi_p(\ell)-F_p(\ell)}
   &\leq K_p(1+\norm{\ell}_\infty)^{p-2},\label{free:eq:sym-error}\\
 \abs{F_p(u)-F_p(w)}
   &\leq K_p\norm{u-w}_\infty
       (1+\norm u_\infty+\norm w_\infty)^{p-2}.\label{free:eq:lip}
\end{align}
\end{lemma}

\begin{proof}
Fix an ordering of the coordinates. Recursively applying
\eqref{free:eq:exact-rec} expresses $\Phi_p$ on the corresponding integer
chamber as a polynomial of degree at most $p-1$. The homogeneous component of degree
$p-1$ is exactly \eqref{free:eq:F-rec}: a summand with $k$ intervals has
degree $k+\sum_j(|I_j|-1)=p-1$. There are only finitely many chambers
and recursive choices for fixed $p$.

On a common wall, any two permitted formulas agree at every integer
point because both count the same pairings. Dilate such a point by
positive integers and compare the highest degree coefficients. The
leading polynomials agree at integer points on the wall, hence at
rational points by homogeneity, and at all real points by polynomial
continuity. This proves consistency at ties and continuity across
walls, including their intersections. Cyclic invariance follows in
the same way from cyclic invariance of the exact count. Each
summand of \eqref{free:eq:F-rec} is nonnegative. Rotating a zero coordinate
to the first position makes every summand vanish when $p\geq2$.

The lower degree polynomial terms give \eqref{free:eq:sym-error}, uniformly
over the finitely many chambers. The gradients of the leading
polynomials are bounded by a constant times
$(1+\norm u_\infty)^{p-2}$. Subdivide the line segment from $u$ to $w$
at ordering walls and apply this bound on each segment. Continuity
at the joining points gives \eqref{free:eq:lip}.
\end{proof}

For example, $F_2(u_1,u_2)=\min(u_1,u_2)$. If the sorted coordinates
of a three-tuple are $r\leq s\leq t$, then
$F_3=rs+r^2/2$. For larger $p$, the cyclic order enters the recursion,
so each recursive call retains that order while rotating a minimum
to the first coordinate.

\subsubsection{Comparing unequal blocks}

Suppose $\sum a_i=\sum b_i=n$. Define the cyclic valley and peak
heights
\begin{equation}\label{free:eq:heights}
 v_0=0,\qquad v_i=\sum_{j=1}^i(a_j-b_j),\qquad
 h_i=v_{i-1}+a_i=v_i+b_i.
\end{equation}
Thus $v_p=v_0$. Put
\[
 m=\min_{0\leq i<p}v_i,\qquad
 M=\max_{0\leq i<p}v_i,\qquad D=M-m.
\]

\begin{lemma}[Valley insertion]\label{free:lem:insertion}
For every balanced pair of nonnegative integer length vectors,
\begin{equation}\label{free:eq:upper-inj}
 P_p(a,b)\leq\Phi_p(h_1-m,\ldots,h_p-m).
\end{equation}
If $\min_i h_i\geq M$, then also
\begin{equation}\label{free:eq:lower-inj}
 \Phi_p(h_1-M,\ldots,h_p-M)\leq P_p(a,b).
\end{equation}
\end{lemma}

\begin{proof}
At the marked gap between $-^{b_i}$ and $+^{a_{i+1}}$, insert
$-^{v_i-m}+^{v_i-m}$ and pair the inserted letters by their nested
opposite-sign matching. Every inserted pair has both endpoints in
that gap, so it crosses no old pair and no insertion at another gap.
The new negative block at peak $h_i$ has length
$b_i+v_i-m=h_i-m$; the positive block there has length
$a_i+v_{i-1}-m=h_i-m$. Retaining every original pair therefore gives
a pairing of the locally symmetric word in \eqref{free:eq:upper-inj}.
The inserted positions are fixed by $a,b$, independently of the
pairing, and their deletion recovers the original pairing. This is
an injection.

The root gap can also be described with fixed linear labels: prepend
$v_p-m$ positive letters and append $v_p-m$ negative letters, then
match these new letters in nested order around the entire old word.
All other insertions are internal. This specifies the root convention
even when the first or last displayed block is empty.

For the lower bound, start with the symmetric lengths $h_i-M$ and
insert $-^{M-v_i}+^{M-v_i}$ at valley $i$. The resulting lengths are
exactly $a,b$, giving another injection. The hypothesis ensures
that all starting lengths are nonnegative. The construction includes
the gap across the cyclic root, with the prescribed root kept fixed.
When some blocks are empty, keep the cyclic order of the marked gaps.
The inserted strings still occupy disjoint consecutive intervals,
and deleting their specified positions remains an inverse on the
image. Thus zero blocks introduce no multiplicity.
\end{proof}

This construction adapts the height-preserving insertion in the proof
of \cite[Proposition~2.16]{KMRS}. Keeping the peaks fixed and inserting
at the marked valleys gives comparisons in both directions, with the
same cyclic block positions. The resulting estimate measures the
error through the range $D$ of the valley heights.

\begin{proposition}\label{free:prop:comparison}
For each fixed $p\geq2$,
\begin{equation}\label{free:eq:comparison}
 \abs{P_p(a,b)-F_p(a)}
 \leq K_p(1+D)(1+n)^{p-2}
\end{equation}
for all balanced nonnegative integer lengths. Moreover, for every
$p\geq1$,
\begin{equation}\label{free:eq:global}
 P_p(a,b)\leq K_p(1+n)^{p-1}.
\end{equation}
Consequently, if $R\to\infty$ and balanced lengths satisfy
$a_i/R\to u_i$ and $b_i/R\to u_i$, where $u\in\R_+^p$, then
\begin{equation}\label{free:eq:sequential}
 R^{-(p-1)}P_p(a,b)\longrightarrow F_p(u).
\end{equation}
\end{proposition}

\begin{proof}
The upper symmetric vector $h-m$ differs from $a$ by coordinates in
$[0,D]$. Since $D\leq n$, its coordinates are at most $2n$.
Combine \eqref{free:eq:upper-inj}, \eqref{free:eq:sym-error}, and
\eqref{free:eq:lip} to obtain the upper side of \eqref{free:eq:comparison}.
When $\min h_i\geq M$, the lower symmetric vector $h-M$ differs
from $a$ by coordinates in $[-D,0]$, giving the other side in the
same way.

If the lower-insertion hypothesis fails, choose $i$ with $h_i<M$.
Then $0\leq a_i<M-v_{i-1}\leq D$. Replace this coordinate of $a$
by zero to obtain $a'$. By Lemma~\ref{free:lem:F}, $F_p(a')=0$ and
$F_p(a)\leq K_pD(1+n)^{p-2}$. Together with $P_p(a,b)\geq0$,
this proves the lower bound in the remaining case.

The upper insertion and the finite polynomial bounds for $\Phi_p$
give \eqref{free:eq:global}. If $a/R,b/R\to u$, then $n=O(R)$ and
$D=o(R)$, so \eqref{free:eq:comparison}, homogeneity, and continuity
give \eqref{free:eq:sequential}. For $p=1$ the balanced count and $F_1$
are both exactly one.
\end{proof}

The comparison also applies on ordering walls and when some
coordinates vanish. In particular, if $u$ ranges over a bounded set and
\[
 \max_i\{\abs{a_i-Ru_i},\abs{b_i-Ru_i}\}\leq L,
\]
then $D\leq2pL$, and for $p\geq2$,
\begin{equation}\label{free:eq:near-diagonal}
 \abs{P_p(a,b)-R^{p-1}F_p(u)}
 \leq K_{p,B}(1+L)(1+R+L)^{p-2},
 \qquad \norm u_\infty\leq B.
\end{equation}
Thus fluctuations of size $O(\sqrt R)$ do not alter the normalized
leading coefficient, even when the limiting tuple lies on a wall.

\subsection{The simplex integrals}

The moment calculation requires the integral of $F_p$ over a simplex.
We evaluate it by averaging the recursion against exponential weights.
The consecutive intervals in the recursion then factor independently,
and their contributions assemble into one generating function.

\begin{proposition}\label{free:prop:integral}
Let
\[
 \Delta_{p-1}=\{u\in\R_+^p:\textstyle\sum_i u_i=1\}.
\]
Equip this simplex with coordinate measure $\dd u_1\cdots\dd u_{p-1}$,
where $u_p=1-\sum_{i<p}u_i$. Then
\begin{equation}\label{free:eq:simplex}
 I_p:=\int_{\Delta_{p-1}}F_p(u)\dd u
       =\frac1{2^{p-1}p!(p-1)!}.
\end{equation}
For $p=1$ the integral means $F_1(1)=1$.
\end{proposition}

\begin{proof}
For $s\geq0$, define
\begin{align*}
 A_p(s)&=\int_{\R_+^p}e^{-\sum v_i}
                F_p(s+v_1,\ldots,s+v_p)\dd v,\\
 B_p(s)&=\int_{\R_+^{p-1}}e^{-\sum_{i=2}^p v_i}
                F_p(s,s+v_2,\ldots,s+v_p)\dd v_2\cdots\dd v_p,
\end{align*}
where $B_1=1$. Common translation preserves every ordering chamber.
On each chamber the integrand $F_p(s+v)$ is polynomial in $s$ of
degree at most $p-1$, with polynomial coefficients in $v$.
There are finitely many such chambers and these coefficients are
integrable against $e^{-\sum v_i}$. Hence $A_p$ is a polynomial of
degree at most $p-1$.

Changing variables gives
\[
 A_p(s)=e^{ps}\int_{[s,\infty)^p}e^{-\sum u_i}F_p(u)\dd u.
\]
Continuity and the polynomial bounds of Lemma~\ref{free:lem:F} justify
differentiation of the lower integration boundaries, for example
by first restricting to a bounded box and then applying exponential
domination. All $p$ boundary integrals agree by cyclic invariance
of $F_p$ and invariance of the weight under cyclic permutations.
Therefore
\begin{equation}\label{free:eq:ODE}
 A_p'(s)=p\bigl(A_p(s)-B_p(s)\bigr).
\end{equation}

Consider the formal power series
$A(z,s)=\sum_{p\geq1}A_p(s)z^p$ in $\R[s][[z]]$.
The first argument in $B_p$ is a minimum. Inserting
\eqref{free:eq:F-rec}, and integrating the disjoint consecutive intervals
separately, gives, for $p\geq2$,
\begin{equation}\label{free:eq:B-compositions}
 B_p(s)=\sum_{k=1}^{p-1}\frac{s^k}{k!}
       \sum_{\substack{m_1+\cdots+m_k=p-1\\m_j\geq1}}
           \prod_{j=1}^k A_{m_j}(s).
\end{equation}
The composition is ordered: each $(m_1,\ldots,m_k)$ specifies exactly
one decomposition of the residual coordinate list. Its factor $1/k!$
comes from \eqref{free:eq:F-rec}, with no additional permutation factor.
Equivalently,
\begin{equation}\label{free:eq:B}
 B_p(s)=[z^{p-1}]\exp(sA(z,s)).
\end{equation}
For $p=1$ this follows from the constant term. Thus
\begin{equation}\label{free:eq:PDE}
 \partial_s A=z\partial_z\bigl(A-z e^{sA}\bigr).
\end{equation}
These are identities in the formal power-series ring: at each degree,
only finitely many compositions contribute.

Equations \eqref{free:eq:ODE}--\eqref{free:eq:B} have at most one solution
with polynomial coefficients $A_p(s)$. At $p=1$, the equation
$A_1'=A_1-1$ has the unique polynomial solution $A_1=1$.
For the induction step, $B_p$ depends only
on $A_1,\ldots,A_{p-1}$. Once those are fixed, two possible $A_p$
differ by a multiple of $e^{ps}$, which cannot be a nonzero
polynomial. This determines the polynomial solution coefficient by
coefficient.

Now let $\widehat A$ be the unique formal series with zero constant
term satisfying
\begin{equation}\label{free:eq:implicit}
 z e^{s\widehat A}=\widehat A-\tfrac12\widehat A^2.
\end{equation}
Lagrange inversion gives
\begin{equation}\label{free:eq:lagrange}
 [z^p]\widehat A
   =\frac1p[w^{p-1}]e^{psw}(1-w/2)^{-p},
\end{equation}
which is polynomial in $s$ of degree at most $p-1$. To check the
equation it solves, set $W=z e^{s\widehat A}$ and
$D=1-\widehat A-sW$. Formal differentiation of
\eqref{free:eq:implicit} gives
\[
 D\,\partial_s\widehat A=W\widehat A,
 \qquad D\,z\partial_z\widehat A=W.
\]
The constant term of $D$ is one. Since
$\widehat A-z e^{s\widehat A}=\widehat A^2/2$, these equations
imply \eqref{free:eq:PDE}. Polynomial uniqueness identifies
$\widehat A=A$.

At $s=0$, formula \eqref{free:eq:lagrange} becomes
\begin{equation}\label{free:eq:J}
 J_p:=\int_{\R_+^p}e^{-\sum u_i}F_p(u)\dd u
      =A_p(0)=\frac{\Cat_{p-1}}{2^{p-1}}.
\end{equation}
Finally, write $u=r\theta$, where $r=\sum u_i$ and
$\theta\in\Delta_{p-1}$. The Jacobian is $r^{p-1}$ and $F_p$
has degree $p-1$, so $J_p=\Gamma(2p-1)I_p$. Using
$\Cat_{p-1}=(2p-2)!/[p!(p-1)!]$ proves \eqref{free:eq:simplex}.
\end{proof}

\subsection{Balanced Poisson averaging and the time integrals}

The following semigroup estimate gives norm convergence of the
covariance integral and a common compact support for its rescaled
spectral measures.

\begin{lemma}\label{free:lem:norm}
For $t\geq0$ and $\delta>0$,
\begin{equation}\label{free:eq:norm}
 \norm{e^{tc}}^2\leq e(t+1)e^{2t},\qquad
 \delta^2\norm{q_\delta}\leq\frac e4+\frac{e\delta}2.
\end{equation}
\end{lemma}

\begin{proof}
Circular Wick moments and the Fuss--Catalan count
\cite[Proposition~1.6]{KMRS} give, for $n,r\geq1$,
\[
 \tau\bigl((c^n c^{*n})^r\bigr)
   =\frac1{nr+1}\binom{(n+1)r}{r}.
\]
Faithfulness of $\tau$, the norm formula for positive operators, and
Stirling's formula imply
\[
 \norm{c^n}^2=\frac{(n+1)^{n+1}}{n^n}\leq e(n+1).
\]
The same upper bound holds for $n=0$. If $K$ is Poisson with mean
$t$, the exponential series and Jensen's inequality give
\[
 \norm{e^{tc}}\leq\sqrt e\,e^t\E\sqrt{K+1}
                  \leq\sqrt{e(t+1)}e^t.
\]
Integrating $e(t+1)e^{-2\delta t}$ proves norm convergence of
\eqref{free:eq:q} and the second bound in \eqref{free:eq:norm}.
\end{proof}

For fixed $\delta>0$, Lemma~\ref{free:lem:norm} makes the $p$-fold product
integral for $\tau(q_\delta^p)$ absolutely integrable in norm.
Expand its exponentials. The circular Wick formula says that each
trace coefficient is exactly $P_p(a,b)$. In particular all scalar
coefficients are nonnegative, so Tonelli's theorem applies as well.
For independent Poisson variables $A_i,B_i$ of mean $t_i$, this gives
the exact identity
\begin{equation}\label{free:eq:poisson}
 \tau(q_\delta^p)=\int_{\R_+^p}e^{-2\delta T}
                   \E P_p(A,B)\dd t,\qquad T=\sum_i t_i.
\end{equation}
The balance condition in this expectation is
$\sum A_i=\sum B_i$, imposed on the totals across all blocks.

\begin{lemma}\label{free:lem:Poisson}
Fix $p\geq1$ and $u_i>0$. Put $U=\sum u_i$, and let all
$A_i,B_i$ be independent Poisson variables of means $Ru_i$.
Then
\begin{equation}\label{free:eq:poisson-limit}
 R^{3/2-p}\E P_p(A,B)
       \longrightarrow\frac{F_p(u)}{\sqrt{4\pi U}}
       \qquad(R\to\infty).
\end{equation}
For $R\geq1$, the left side is bounded by
\begin{equation}\label{free:eq:dom}
 \begin{cases}
 K U^{-1/2},&p=1,\\
 K_p(1+U)^{p-3/2},&p\geq2.
 \end{cases}
\end{equation}
\end{lemma}

\begin{proof}
Let $E_R=\{\sum A_i=\sum B_i\}$ and $\lambda=RU$. The two totals
are independent Poisson variables of mean $\lambda$, so Fourier
inversion gives
\begin{equation}\label{free:eq:balance}
 \Prob(E_R)=\frac1{2\pi}\int_{-\pi}^{\pi}
          e^{2\lambda(\cos\theta-1)}\dd\theta
       \sim\frac1{\sqrt{4\pi RU}}.
\end{equation}
To obtain the asymptotic, rescale $\theta=w/\sqrt\lambda$.
The integrand tends to $e^{-w^2}$ and is bounded by $e^{-cw^2}$
on its expanding domain because $1-\cos\theta\geq c\theta^2$
for $\abs\theta\leq\pi$. Dominated convergence proves the
asymptotic; the same integral gives the bound
$\Prob(E_R)\leq K(1+\lambda)^{-1/2}$.

Let $G_R$ be the event that each of the $2p$ variables is within
$R^{3/4}$ of its mean. For fixed $u$, the exponential generating
function of a Poisson variable gives the Chernoff bound
\begin{equation}\label{free:eq:chernoff}
 \Prob(G_R^c)\leq K_u e^{-c_u\sqrt R}.
\end{equation}
On $E_R\cap G_R$, \eqref{free:eq:near-diagonal} gives, for $p\geq2$,
\[
 \sup\abs{R^{-(p-1)}P_p(A,B)-F_p(u)}=O_{p,u}(R^{-1/4}).
\]
For $p=1$ the difference is zero. On the exceptional event, use
\eqref{free:eq:global}, Cauchy--Schwarz, and the fixed-order Poisson
moment bound to obtain
\begin{equation}\label{free:eq:bad}
 \E\bigl[P_p(A,B)\one_{G_R^c}\bigr]
 \leq K_{p,u}R^{p-1}e^{-c'_u\sqrt R}.
\end{equation}
Indeed $P_p$ vanishes off $E_R$, and on $E_R$ it is at most
$K_p(1+\sum A_i)^{p-1}$; the second moment of this upper bound is
$O_{p,u}(R^{2p-2})$. More explicitly, for $p\geq2$ these estimates give
\begin{align}
 &\abs{\E P_p(A,B)-R^{p-1}F_p(u)\Prob(E_R)}
 \notag\\
 &\hspace{1em}\leq
 K_{p,u}R^{p-1-1/4}\Prob(E_R)
       +K_{p,u}R^{p-1}e^{-c'_{p,u}\sqrt R}.
 \label{free:eq:balanced-error}
\end{align}
The first term retains the probability of total balance. Multiplication
by $R^{3/2-p}$ and \eqref{free:eq:balance} make both errors tend to zero.
For $p=1$, $\E P_1=\Prob(E_R)$ exactly. This proves
\eqref{free:eq:poisson-limit}.

To obtain a bound uniform over all time vectors, let $K_1,K_2$
be independent Poisson variables of mean $\lambda$ and put $d=p-1$.
Stirling's bound at the mode gives
\[
 \sup_k\Prob(K_2=k)\leq K(1+\lambda)^{-1/2},
 \qquad \E(K_1+1)^d\leq K_d(1+\lambda)^d.
\]
Consequently
\begin{align*}
 \E\bigl[(K_1+1)^d\one_{\{K_1=K_2\}}\bigr]
 &=\sum_{k\geq0}(k+1)^d\Prob(K_1=k)\Prob(K_2=k)\\
 &\leq K_d(1+\lambda)^{d-1/2}.
\end{align*}
Apply \eqref{free:eq:global} with these two totals. This bounds the
left side of \eqref{free:eq:poisson-limit} by
$K_p(U+R^{-1})^{p-3/2}$, which implies \eqref{free:eq:dom}.
\end{proof}

Conditional on $E_R$ and its common total, the two block vectors
are independent multinomial vectors with cell probabilities $u_i/U$.
The comparison estimate absorbs their fluctuations into a lower-order
error. Thus the leading coefficient survives the Poisson averaging,
multiplied by the probability of total balance, whose order is
$R^{-1/2}$.

\begin{proof}[Proof of the moment assertion in Theorem~\ref{free:thm:main}]
Set $R=\delta^{-1}$ and $t_i=Ru_i$ in \eqref{free:eq:poisson}. Then
\begin{equation}\label{free:eq:scaled-int}
 \delta^{2p-3/2}\tau(q_\delta^p)
  =\int_{\R_+^p}e^{-2U}R^{3/2-p}\E P_p(A,B)\dd u.
\end{equation}
The functions in \eqref{free:eq:dom}, multiplied by $e^{-2U}$, are
integrable over this orthant. Indeed, radial integration introduces
the Jacobian $U^{p-1}$; when $p=1$, the singularity $U^{-1/2}$ is
integrable at the origin. The same bounds control neighborhoods of
all boundary faces and the tail $U\to\infty$. Pointwise convergence
on the open orthant therefore suffices, since its boundary has
Lebesgue measure zero. The estimates already include Poisson samples
with zero blocks. Dominated convergence gives
\begin{align}
 C_p&=\frac1{\sqrt{4\pi}}\int_{\R_+^p}
              e^{-2U}U^{-1/2}F_p(u)\dd u\notag\\
 &=\frac{\Gamma(2p-\tfrac32)}{2^{2p-1/2}\sqrt\pi}\,I_p.
 \label{free:eq:C-integral}
\end{align}
Insert \eqref{free:eq:simplex} to obtain \eqref{free:eq:moments}.
\end{proof}

The first moment also has the exact expression obtained from
\eqref{free:eq:balance} and \eqref{free:eq:poisson}:
\[
 \tau(q_\delta)=\frac1{4\pi}\int_{-\pi}^{\pi}
                       \frac{\dd\theta}{1+\delta-\cos\theta}
              =\frac1{2\sqrt{(1+\delta)^2-1}}.
\]
This agrees with $C_1$ in \eqref{free:eq:checks}.

\subsection{Identifying the spectral law}

\begin{proof}[Proof of the remaining assertions of Theorem~\ref{free:thm:main}]
Lemma~\ref{free:lem:norm} puts all $\nu_\delta$, for $0<\delta\leq1$,
in the fixed compact interval $[0,3e/4]$. Their masses converge to
$C_1$. Their moments satisfy, for each fixed $k\geq0$,
\begin{equation}\label{free:eq:nu-moments}
 \int x^k\nu_\delta(\dd x)
    =\delta^{2k+1/2}\tau(q_\delta^{k+1})
    \longrightarrow C_{k+1}.
\end{equation}
Compactness gives weakly convergent subsequences. Their limits agree
on polynomials by \eqref{free:eq:nu-moments}, and hence on all continuous
functions by uniform polynomial approximation. The limit is therefore
unique, and the full sequence converges weakly.

The moment constants identify this limit. With $(a)_k$ denoting
the rising factorial, gamma duplication gives
\begin{equation}\label{free:eq:beta-moments}
 \frac{C_{k+1}}{C_1}
 =2^{-k}\frac{(1/4)_k(3/4)_k}{(1)_k(2)_k}
 =\E\left(\frac{UV}{2}\right)^k.
\end{equation}
The last equality is the elementary beta moment formula and
independence. Compact moment uniqueness gives \eqref{free:eq:betalaw}.
The product $X$ has no atoms, has support $[0,1/2]$, and has a
positive density on $(0,1/2)$.

The definition of $\nu_\delta$ gives the exact identity
\begin{equation}\label{free:eq:tail-identity}
 \delta^{-3/2}\mu_\delta((x\delta^{-2},\infty))
       =\int_{(x,\infty)}y^{-1}\nu_\delta(\dd y),\qquad x>0.
\end{equation}
For fixed $x$, the test function on the right is bounded and has
only one discontinuity, at $x$. The limiting measure has no atom
there. Weak convergence proves \eqref{free:eq:tail} for every $x>0$,
including $x=1/2$. The beta representation gives the stated
continuity, positivity, and support of $H$.

Finally fix $0<a<b<\infty$. Partition $[a,b]$ into finitely many
intervals on each of which the oscillation of the continuous
function $H$ is less than $\varepsilon$. Pointwise convergence at
the endpoints, and monotonicity of both spectral tails and $H$,
bound the difference uniformly by $2\varepsilon$ for small
$\delta$. This proves local uniform convergence.
\end{proof}

\subsubsection{An integral density and the two endpoints}

Let $\rho$ be the density of the finite measure $\nu_*$. The two
beta densities have normalizing constants $1/(\pi\sqrt2)$ and
$2\sqrt2/\pi$. The product-density formula and the rescaling
$X=UV/2$ give, for $0<x<1/2$,
\begin{equation}\label{free:eq:density}
 \rho(x)=\frac{2^{-1/4}}{\pi^2}x^{-3/4}
       \int_{2x}^1 v^{-1/2}(1-v)^{1/4}
                         (1-2x/v)^{-1/4}\dd v.
\end{equation}
In particular,
\begin{equation}\label{free:eq:H-density}
 H(x)=\int_x^{1/2}\frac{\rho(y)}y\dd y\quad(0<x<1/2).
\end{equation}

\begin{proof}[Proof of Corollary~\ref{free:cor:ends}]
Put $z=2x$. On $v\geq2z$ the last factor in the integral of
\eqref{free:eq:density} is bounded, so dominated convergence applies
as $z\downarrow0$. The remaining interval $z\leq v\leq2z$
contributes $O(\sqrt z)$: substitute $v=z(1+s)$ and bound the
resulting integrand by a constant times $s^{-1/4}$. Hence the
integral converges to $\mathrm B(1/2,5/4)$. Simplifying with
$\Gamma(1/4)\Gamma(3/4)=\pi\sqrt2$ gives
\[
 \rho(x)\sim\frac{2^{1/4}}{3\sqrt\pi\,\Gamma(3/4)^2}x^{-3/4}.
\]
Integrating \eqref{free:eq:H-density} proves \eqref{free:eq:small}.
For $x_r=H^{-1}(r)$, we have $x_r\downarrow0$ as $r\to\infty$.
Thus $r x_r^{3/4}\to c_0$, which proves the inverse asymptotic.

For the upper endpoint, rewrite the integrand in
\eqref{free:eq:density} as
$v^{-1/4}(1-v)^{1/4}(v-z)^{-1/4}$ and put
$v=z+(1-z)s$. The integral becomes
\[
 (1-z)\int_0^1[z+(1-z)s]^{-1/4}
                       (1-s)^{1/4}s^{-1/4}\dd s
 \sim(1-z)\mathrm B(3/4,5/4).
\]
The integrable beta weight justifies this limit. The constants in
\eqref{free:eq:density} now give
$\rho(x)\sim\pi^{-1}(1/2-x)$.
Since $1/y\to2$ at this endpoint, \eqref{free:eq:H-density} yields
\eqref{free:eq:edge}.
\end{proof}

\subsubsection{The relation to the discrete-time profile}

The formal discrete-time calculation in \cite[Eq.~(62)]{ZV} gives a
critical tail profile which, after weighting by the eigenvalue and
normalizing the mass, has density
\begin{equation}\label{free:eq:discrete-energy}
 \psi_d(x)=\frac1\pi\sqrt{\frac{2-x}{x}}\,
                  \one_{\{0<x<2\}}.
\end{equation}
In that paper $g$ is the variance parameter of the interaction matrix,
and the spectral variable is $x=(1-g)^2\lambda$.
The following identity relates this explicit density to the
probability measure $\nu_*/C_1$.

\begin{proposition}\label{free:prop:discrete-profile}
Let $D$ have density \eqref{free:eq:discrete-energy}, and let
$A\sim\Beta(1/2,1/2)$ be independent of $D$. Then
\begin{equation}\label{free:eq:arcsine-product}
 \frac{\nu_*}{C_1}=\Law\left(\frac{DA^2}{4}\right).
\end{equation}
\end{proposition}
\begin{proof}
On $(0,1)$, the random variable $B:=D/2$ has density
\[
 \frac{2}{\pi}b^{-1/2}(1-b)^{1/2}.
\]
Thus $B\sim\Beta(1/2,3/2)$. For every integer $k\ge0$,
independence and the duplication formula give
\[
 \E[(B A^2)^k]
 =\frac{(1/2)_k}{(2)_k}\frac{(1/2)_{2k}}{(1)_{2k}}
 =\frac{(1/4)_k(3/4)_k}{(1)_k(2)_k}
 =\E[(UV)^k].
\]
Both products take values in $[0,1]$, where moments determine the
law. Hence $UV$ and $BA^2$ have the same distribution, proving
\eqref{free:eq:arcsine-product} by \eqref{free:eq:betalaw}.
\end{proof}

The factor $1/4$ aligns the two stability parameters. For $0<g<1$, set
$a=1-\sqrt g$, the gap between the limiting interaction radius and
one. Since $(1-g)/a=1+\sqrt g\to2$, the discrete profile on the
$a^2\lambda$ scale is $D/4$. For the continuous model, with the
standard circular element $c$ used here, write
\[
 \Sigma_c(g)=\int_0^\infty e^{-2t}
                      e^{\sqrt g\,tc}e^{\sqrt g\,tc^*}\dd t.
\]
If $g=(1+\delta)^{-2}$, the change of variables $t=(1+\delta)s$
gives
\[
 \Sigma_c(g)=(1+\delta)q_\delta,\qquad
 a^2\Sigma_c(g)=\frac{\delta^2q_\delta}{1+\delta}.
\]
The multiplicative factor in the covariance cancels upon normalizing
the energy measure. More explicitly, for every bounded continuous $f$,
\[
 \frac{\tau[\Sigma_c(g)f(a^2\Sigma_c(g))]}{\tau\Sigma_c(g)}
 =\frac{\tau[q_\delta f(\delta^2q_\delta/(1+\delta))]}
        {\tau q_\delta}
 \longrightarrow\E f(X).
\]
Indeed, the weighted circular measures converge to $\nu_*$, their
masses converge to $C_1>0$, and the rescaling $x\mapsto x/(1+\delta)$
tends uniformly to the identity on their common compact support.
Thus, on the common gap scale, the limiting normalized continuous
energy profile is the product of the discrete profile $D/4$ and $A^2$.
Both limiting energy profiles have upper endpoint $1/2$.

The arcsine factor can also be read directly from the moment
calculation. Equations~\eqref{free:eq:J} and
\eqref{free:eq:C-integral} give
\[
 \frac{C_{k+1}}{C_1}
 =\frac{J_{k+1}}{4^k}
       \frac{\Gamma(2k+1/2)}{\sqrt\pi\,\Gamma(2k+1)}
 =\E[(D/4)^k]\,\E A^{2k}.
\]
At each fixed moment order, the global Poisson balance contributes
the factor $U^{-1/2}$ in \eqref{free:eq:C-integral}. Its effect on
the radial integral is exactly the second factor in this identity.

\subsection{Finite observation times}

For $T<\infty$ let
\[
 q_{\delta,T}=\int_0^T e^{-2(1+\delta)t}e^{tc}e^{tc^*}\dd t.
\]
The cutoff $T=L/\delta$ keeps a fixed multiple of the damping time.
Its limiting spectral measure will provide the link to the finite
matrix estimates.

\begin{lemma}\label{lem:finite-free}
For each fixed $L>0$ and integer $p\ge1$,
\begin{equation}\label{eq:Dp}
 \delta^{2p-3/2}\tau(q_{\delta,L/\delta}^{p})
 \longrightarrow
 D_p(L):=\frac1{\sqrt{4\pi}}\int_{[0,L]^p}
       e^{-2U}U^{-1/2}F_p(u)\dd u .
\end{equation}
There is a unique finite measure $\nu_L$ supported on $[0,3e/4]$
whose moments are
\begin{equation}\label{eq:nuL-moments}
 \int x^k\,\nu_L(\dd x)=D_{k+1}(L),\qquad k\ge0.
\end{equation}
The weighted spectral measures of $q_{\delta,L/\delta}$ converge
weakly to $\nu_L$ as $\delta\downarrow0$, and
$\nu_L\Rightarrow\nu_*$ as $L\to\infty$.
\end{lemma}
\begin{proof}
Expand the exponentials and apply the circular Wick formula.
Nonnegativity of the scalar coefficients permits Tonelli's theorem.
With $t_i=u_i/\delta$ and $R=\delta^{-1}$, the scaled moment is
\[
 \int_{[0,L]^p}e^{-2U}R^{3/2-p}\E P_p(A,B)\dd u.
\]
The Poisson limit \eqref{free:eq:poisson-limit} and its integrable
majorant \eqref{free:eq:dom} give \eqref{eq:Dp} by dominated convergence,
including neighborhoods of the boundary faces and the origin.

Positivity gives $q_{\delta,L/\delta}\le q_\delta$.
Equation~\eqref{free:eq:norm} therefore puts the weighted measures
in $[0,3e/4]$ for $0<\delta\le1$. Their masses converge by the
case $p=1$, and all their moments converge by \eqref{eq:Dp}.
Compact moment determinacy proves existence, uniqueness and
full-sequence weak convergence to $\nu_L$.

For the full covariance, the Poisson representation has integration
domain $[0,\infty)^p$. The majorant in \eqref{free:eq:dom}, multiplied
by $e^{-2U}$, remains integrable there. Thus dominated convergence
and \eqref{free:eq:moments} identify
\[
 C_p=\frac1{\sqrt{4\pi}}\int_{[0,\infty)^p}
             e^{-2U}U^{-1/2}F_p(u)\dd u .
\]
With this full-orthant integral identified, monotone convergence gives
$D_p(L)\uparrow C_p$ as $L\to\infty$. The common compact support and
moment convergence then imply $\nu_L\Rightarrow\nu_*$. Each moment
order remains fixed: the limit $\delta\downarrow0$ is taken first,
followed by $L\to\infty$.
\end{proof}

\section{Bounded symmetric entries}\label{sec:comparison}

Let $X_N=N^{-1/2}(\xi_{ij})$, where the entries are independent
copies of a real random variable $\xi$ satisfying
\[
 \E\xi=0,\qquad \E\xi^2=1,\qquad |\xi|\le M.
\]
In particular $M\ge1$. We begin with the trace terms in which every
matrix entry has even multiplicity. These terms admit a comparison
with Gaussian pairings, whose cost is controlled by the topological
deficit and by collisions among vertex labels. For a symmetric entry
law they exhaust the expectation, so the comparison already gives
the required trace estimates on growing time intervals. The next
section treats the odd multiplicities.

For a real matrix $A$, set
\[
 Q_{\delta,T}(A)=\int_0^T e^{-2(1+\delta)t}
                  e^{tA}e^{tA^{\mathsf T}}\dd t,
 \qquad F_{p,\delta,T}(A)=\tau_N Q_{\delta,T}(A)^p.
\]
Throughout the comparison, a sign word denotes a product of $A$ and
$A^{\mathsf T}$.
In its cyclic trace expansion, a positive letter at position $j$
uses the ordered entry $(i_j,i_{j+1})$, whereas a negative letter
uses $(i_{j+1},i_j)$. We call these ordered pairs the physical
entries of the index assignment.

\begin{lemma}[The collision factor]\label{lem:bounded-collisions}
Give $v$ distinct formal vertices independent uniform labels in
$\{1,\ldots,N\}$. Let $c$ be $v$ minus the number of distinct
labels. For every $a\ge1$,
\begin{equation}\label{eq:bounded-collision-mgf}
 \E a^c\le\prod_{j=0}^{v-1}\left(1+\frac{(a-1)j}{N}\right)
 \le\exp\left(\frac{(a-1)v(v-1)}{2N}\right).
\end{equation}
\end{lemma}
\begin{proof}
Expose the labels in a fixed order. The next label creates one
collision precisely when it is among the labels already seen.
After $j$ labels this conditional probability is at most $j/N$.
The conditional expectation of the factor contributed by the next
label is therefore at most $1+(a-1)j/N$. Iteration proves the first
inequality, and $1+x\le e^x$ proves the second. The argument also
applies when $v>N$, since the upper bound on the conditional
probability need not be less than one.
\end{proof}

For a sign word $w$ of even length $2n$, write $\mathcal E_N(w)$
for the part of $\E\tau_Nw(X_N)$ coming from index assignments
whose physical-entry multiplicities are all even. For two words
$w_1,w_2$ of total length $2n$, write $\mathcal V_N(w_1,w_2)$
for the sum of the absolute values of their normalized monomial
covariances over such assignments. The multiplicities in the
second definition are taken in the union of the two words.
Let $\mathcal P_1(w)$ denote the real Wick pairings of the first
word, and let $\mathcal P_2^{\mathrm c}(w_1,w_2)$ denote those
pairings of the two words that contain a pair joining the faces.
These are exactly the connected two-face maps. The deficit $h(\pi)$
is the one in \eqref{gauss:eq:real-weight}.

\begin{lemma}[Even-multiplicity comparison]\label{lem:bounded-even}
For a nonempty word of length $2n$,
\begin{equation}\label{eq:bounded-even-mean}
 \mathcal E_N(w)
 \le e^{C_M(n+1)^2/N}
       \sum_{\pi\in\mathcal P_1(w)}M^{2h(\pi)}N^{-h(\pi)}.
\end{equation}
For two nonempty words of total length $2n$,
\begin{equation}\label{eq:bounded-even-var}
 \mathcal V_N(w_1,w_2)
 \le 3M^2e^{C_M(n+1)^2/N}
       \sum_{\pi\in\mathcal P_2^{\mathrm c}(w_1,w_2)}
                   M^{2h(\pi)}N^{-2-h(\pi)}.
\end{equation}
Both statements hold without assuming symmetry of the entry law.
The empty word has expectation one and zero covariance.
\end{lemma}
\begin{proof}
Fix an index assignment with physical entries $e$ of even
multiplicities $m_e$. Put $\mu_m=\E\xi^m$ and $\mu_0=1$. Its entry moment is
$J=\prod_e\mu_{m_e}\ge0$. Refining each physical-entry class into
pairs gives exactly
\[
 R=\prod_e(m_e-1)!!
\]
real Wick pairings compatible with this assignment. Give each
refinement the weight
\[
 W=\prod_e\frac{\mu_{m_e}}{(m_e-1)!!}.
\]
Their total weight is $RW=J$.

Fix one such pairing with $f\in\{1,2\}$ faces, connected when
$f=2$. It has $n$ edges and
\[
 V'=n+2-f-h
\]
formal index vertices. Every compatible cyclic index assignment
is uniquely a labeling of these vertices; distinct formal vertices
are allowed to receive the same label. Suppose there are $V$
distinct labels and put $c=V'-V$. If $q$ physical ordered entries
remain after labeling, their underlying multigraph is connected,
so $q\ge V-1$. Consequently
\[
 n-q\le n-V+1=f-1+h+c.
\]
Since $\mu_m\le M^{m-2}$ for even $m\ge2$,
\begin{equation}\label{eq:bounded-refinement-weight}
 W\le M^{2(n-q)}\le M^{2(f-1)}M^{2h}M^{2c}.
\end{equation}
Summing over all $N^{V'}$ labelings and applying
Lemma~\ref{lem:bounded-collisions}, with $a=M^2$ and
$V'\le n+1$, bounds the contribution of this pairing by
\[
 M^{2(f-1)}M^{2h}e^{C_M(n+1)^2/N}N^{V'-n-f}
 =M^{2(f-1)}M^{2h}e^{C_M(n+1)^2/N}N^{2-2f-h}.
\]
For one face, summing all refinements proves
\eqref{eq:bounded-even-mean}.

For the covariance, fix two index assignments. If their physical-entry
sets are disjoint, their covariance vanishes. Otherwise write
$a_e,b_e$ for the multiplicities in the two faces, so $m_e=a_e+b_e$
is even for every $e$. H\"older's inequality gives
\[
 |\E\xi^{a_e}\,\E\xi^{b_e}|
 \le \E|\xi|^{a_e}\,\E|\xi|^{b_e}
 \le \E|\xi|^{a_e+b_e}=\mu_{m_e}.
\]
Independence now bounds both the joint moment and the product of the
separate moments by $J$, and hence bounds the absolute covariance
of the two unnormalized entry monomials by $2J$.

At a shared physical entry, both $a_e$ and $b_e$ are positive.
If they are odd, every pairing refinement has a cross-face pair.
If they are even, the proportion of refinements at that entry
without a cross-face pair is
\[
 \frac{(a_e-1)!!(b_e-1)!!}{(a_e+b_e-1)!!}\le\frac13.
\]
To see the inequality, put $a_e=2a$, $b_e=2b$ with $a,b\ge1$.
The ratio decreases when either $a$ or $b$ increases, and equals
$1/3$ at $a=b=1$. Thus at least $2R/3$ of all refinements connect
the two faces. If their number is $R_{\mathrm c}$, then
\[
 |\operatorname{Cov}(U,V)|\le2J
    \le3R_{\mathrm c}W.
\]
Sum this inequality over the index assignments, and then sum the
connected refinements using \eqref{eq:bounded-refinement-weight}
with $f=2$. This proves \eqref{eq:bounded-even-var}.
\end{proof}

\begin{lemma}[Planar terms]\label{lem:bounded-planar}
For a sign word $w$ of length $2n$, let $P(w)$ be its number of noncrossing
opposite-sign pairings. Then
\begin{equation}\label{eq:bounded-planar-lower}
 \mathcal E_N(w)
 \ge\frac{(N)_{n+1}}{N^{n+1}}P(w)
 \ge\left(1-\frac{n(n+1)}{2N}\right)P(w).
\end{equation}
Here $(N)_k=0$ for $k>N$. The empty word has $P(w)=1$,
and odd-length words have $\mathcal E_N(w)=0$.
\end{lemma}
\begin{proof}
Every even physical-entry moment is nonnegative. A noncrossing opposite-sign pairing
has a tree as its index graph, with $n$ edges and $n+1$ vertices.
Assign distinct labels to its vertices. Each physical ordered
entry then occurs exactly twice and contributes its variance,
which is one before the matrix normalization. There are
$(N)_{n+1}$ such assignments, each of normalized weight $N^{-n-1}$.
They are disjoint for different pairings: equality of physical
ordered entries identifies the unique paired positions, and then
the vertex labels. Indeed a tree has neither loops nor parallel
edges, and its vertex labeling is injective. This proves the first
inequality. The probability that $n+1$ independent uniform labels
are distinct is at least $1-n(n+1)/(2N)$ by the union bound,
which proves the second. It remains valid when $n\ge N$ because
its right-hand factor is then nonpositive. The case $n=0$ is exact.
\end{proof}

\begin{theorem}[Growing observation times]\label{prop:bounded-traces}
Suppose the distribution of $\xi$ is symmetric. For each fixed
integer $p\ge1$ there are positive constants
$A_{p,M},B_{p,M},C_{p,M},c_{p,M}$ such that the following holds.
Let $\delta\ge0$, $T\ge1$, and $K=\lceil A_{p,M}T\rceil$.
If
\[
 B_{p,M}(K+1)^2/N\le\tfrac12,\qquad T\le c_{p,M}N,
\]
then
\begin{align}
 \left|\E\tau_NQ_{\delta,T}(X_N)^p-\tau(q_{\delta,T}^p)\right|
  &\le C_{p,M}N^{-1}T^{2p+1/2}+\zeta_{p,M},
       \label{eq:bounded-trace-mean}\\
 \operatorname{Var}\bigl(\tau_NQ_{\delta,T}(X_N)^p\bigr)
  &\le C_{p,M}N^{-2}T^{4p+1/2}+\zeta_{p,M},
       \label{eq:bounded-trace-var}
\end{align}
where $\zeta_{p,M}=C_{p,M}(e^{-c_{p,M}T}+e^{-c_{p,M}N})$.
\end{theorem}
\begin{proof}
We first truncate the exponential series, compare the resulting
polynomial Gramians, and then estimate the truncation error. Set
\[
 P_t=e^{-(1+\delta)t}\sum_{j=0}^K\frac{t^jX_N^j}{j!},
 \qquad A_N=\int_0^TP_tP_t^{\mathsf T}\dd t.
\]
Write $A_c$ for the corresponding polynomial in the circular
element. Every scalar coefficient in the entry expansion of
$\tau_NA_N^p$ is nonnegative. Because the entry law is symmetric,
its nonzero expectation terms are exactly the even-multiplicity
terms in Lemma~\ref{lem:bounded-even}.

Each word has at most $2p$ sign changes and at most $pK$ pairs.
After increasing $B_{p,M}$, the exponential collision factor in
\eqref{eq:bounded-even-mean} is bounded by a constant. Its excess
over one is at most $C_{p,M}(n+1)^2/N$. For a planar pairing this
excess contributes at most
\[
 \frac{C_{p,M}}N(K+1)^{p+1}
       \one_{\{a_1+\cdots+a_p=b_1+\cdots+b_p\}},
\]
by the circular count \eqref{free:eq:global}. The nonplanar
contribution is controlled by Lemma~\ref{gauss:lem:real-count}.
The factor $M^{2h}$ changes the constant in its geometric
deficit sum, whose ratio is at most $B_{p,M}(K+1)^2/N$.
As in \eqref{gauss:eq:balance-union}, the resulting per-word
upper errors are bounded by
\[
 \frac{C_{p,M}}N(K+1)^{p+1}
       \sum_{\eta\in\{1,-1\}^{2p}}\one_{\{\eta\cdot a=0\}}
 \quad\hbox{and}\quad
 \frac{C_{p,M}}N(K+1)^p
\]
for the bipartite and nonbipartite block graphs, respectively.
Here $a$ denotes all displayed block lengths. The lower error
follows from Lemma~\ref{lem:bounded-planar} and is bounded by
the first expression with the original positive-negative balance.

For a time vector $t$ put $u=\sum_i t_i$. Multiplication by
$e^{-2u}$ turns the scalar coefficients into probabilities of
independent Poisson block lengths with means
$t_1,t_1,\ldots,t_p,t_p$. The coefficient sums are nonnegative,
so the truncated sums can be bounded by the full Poisson sums
after the finite deficit sum has been estimated.
Lemma~\ref{gauss:lem:real-poisson} bounds every displayed balance
probability by $C_p(1+u)^{-1/2}$. Since
\[
 \int_{[0,T]^m}(1+\textstyle\sum_i t_i)^{-1/2}\dd t
       \le C_mT^{m-1/2},\qquad T\ge1,
\]
and $K+1\le C_{p,M}T$, the polynomial mean error is at most
$C_{p,M}N^{-1}T^{2p+1/2}$.

For the variance, sum the absolute monomial covariances with their
nonnegative coefficients. Symmetry eliminates every term having an
odd multiplicity in the union of the two words, and independence
eliminates terms with disjoint physical-entry sets. Apply
\eqref{eq:bounded-even-var} to the remaining terms. Connected
two-face maps have weights $N^{-2-h}$, at most
$4p$ sign changes, and at most $2pK$ pairs. The collision factor
is bounded uniformly, and the factors $3M^2$ and $M^{2h}$ are
absorbed into the constants and the finite geometric sum. The
per-word bounds from Lemma~\ref{gauss:lem:real-count} are therefore
\[
 C_{p,M}N^{-2}(K+1)^{2p+1}
 \quad\hbox{and}\quad C_{p,M}N^{-2}(K+1)^{2p}
\]
for the bipartite and nonbipartite classes. The first is multiplied
by the union of the $4p$-block balance indicators. Poisson summation
with $m=2p$ gives
$\operatorname{Var}(\tau_NA_N^p)\le C_{p,M}N^{-2}T^{4p+1/2}$.
Words of degree zero contribute no covariance. The Gaussian counting
and coefficient lemmas also include pure faces and vanishing
displayed blocks.

It remains to pass from polynomials to the exponential. Hoeffding's
lemma and independence give, for real unit vectors $u,v$,
\[
 \E e^{t u^{\mathsf T}X_Nv}\le e^{M^2t^2/(2N)}.
\]
Two $1/8$-nets therefore yield
\[
 \Prob(\|X_N\|>x)
 \le2\,17^{2N}e^{-9Nx^2/(32M^2)}
 \le e^{-Nx^2/(8M^2)},\qquad x\ge10M.
\]
Put $R=10M$ and choose $A_{p,M}=100(p+1)(R+1)$.
On $\|X_N\|\le R$, the exponential remainder satisfies
\[
 \sup_{0\le t\le T}
 \left\|e^{-(1+\delta)t}e^{tX_N}-P_t\right\|
 \le e^{RT}\left(\frac{eRT}{K+1}\right)^{K+1}
 \le e^{RT-4(K+1)}.
\]
The norm of either semigroup is at most $e^{RT}$. Telescoping
first the Gramian product and then its $p$th power bounds the
normalized trace error by
$2pT^p e^{2pRT-4(K+1)}$, which decays exponentially in $T$.
The circular estimate is identical because $\|c\|=2\le R$.

On the complementary event, put $R_N=\|X_N\|$. Both Gramians
have norm at most $Te^{2TR_N}$, so the squared trace error is at
most $4T^{2p}e^{4pTR_N}$. If $c_{p,M}$ is small enough, integration
against the displayed norm tail bounds its expectation on
$\{R_N>R\}$ by $C_{p,M}e^{-c_{p,M}N}$. Indeed
$4pTx\le Nx^2/(16M^2)$ for all $x\ge R$ once
$T\le c_{p,M}N$, and the remaining Gaussian tail absorbs the
polynomial factor in $N$. Thus, after adjusting the constants,
$\zeta_{p,M}$ bounds both the $L^1$ approximation error and its
squared $L^2$ error. The first bound transfers the mean estimate to
the exponential Gramian; the second, together with
\[
 \operatorname{Var}Z\le2\operatorname{Var}W+2\E|Z-W|^2
\]
transfers the variance estimate.
\end{proof}

\begin{corollary}\label{cor:fnm-fixed-L}
Suppose the entry law is symmetric. If $\delta_N\to0$ and
$\sqrt N\delta_N\to\infty$, then for every fixed integer $p\ge1$
and every $L>0$,
\[
 \delta_N^{2p-3/2}\tau_NQ_{\delta_N,L/\delta_N}(X_N)^p
       \xrightarrow{\Prob}D_p(L).
\]
\end{corollary}
\begin{proof}
Take $T=L/\delta_N$ in Theorem~\ref{prop:bounded-traces}. Its
conditions hold eventually. After normalization, the mean error
is $O_{p,M,L}(N^{-1}\delta_N^{-2})$, and the variance is
$O_{p,M,L}(N^{-2}\delta_N^{-7/2})$. Writing
$a_N=\sqrt N\delta_N\to\infty$, these are respectively
$O(a_N^{-2})$ and $O(N^{-1/4}a_N^{-7/2})$; the exponential
remainders also vanish. The circular finite-time limit and
Chebyshev's inequality give the result.
\end{proof}

\section{Bounded entries without symmetry}
\label{sec:nonsymmetric}

For a nonsymmetric entry law, the even terms retain the estimates of
the preceding section. The new terms contain physical entries of odd
multiplicity. Closure of the trace forces these entries to form an
even-degree subgraph, producing a positive vertex deficit in addition
to a factor $N^{-1/2}$ for each odd entry. The count in
Appendix~\ref{app:physical-reconstruction} preserves this gain at
growing word lengths, with the original block lengths prescribed.
Its connected two-face version gives the corresponding covariance
bound.

\subsection{The odd terms on a growing time interval}

\begin{proposition}[Odd contributions]
\label{prop:odd-time-sums}
Let $X_N=N^{-1/2}(\xi_{ij})$, where the entries are independent copies
of a real random variable satisfying
\[
 \E\xi=0,\qquad \E\xi^2=1,\qquad |\xi|\le M.
\]
Fix an integer $p\ge1$. For $T\ge1$, $\delta\ge0$ and an integer
$K\ge1$, put
\[
 P_{K,t}=e^{-(1+\delta)t}\sum_{j=0}^K\frac{t^jX_N^j}{j!},
 \qquad A_{N,\delta,T,K}=\int_0^T P_{K,t}P_{K,t}^{\mathsf T}\dd t.
\]
Let $R_1$ be the sum of the absolute values of the terms in the index
expansion of $\E\tau_N A_{N,\delta,T,K}^p$ with at least one odd
physical-entry multiplicity. Let $R_2$ be the corresponding absolute
sum of monomial covariances in
$\operatorname{Var}(\tau_N A_{N,\delta,T,K}^p)$, where multiplicities
are counted in the union of the two monomials. There are constants
$B_{p,M},C_{p,M}$ such that, if
\begin{equation}\label{odd:eq:uniform-time-cutoff}
 \frac{B_{p,M}(1+4pK)^2}{N}\le\frac12,
 \qquad
 \frac{B_{p,M}(1+4pK)}{\sqrt N}\le\frac12,
\end{equation}
then
\begin{align}
 R_1&\le C_{p,M}N^{-3/2}T^{2p+1},
 \label{odd:eq:uniform-time-mean}\\
 R_2&\le C_{p,M}N^{-5/2}T^{4p+2}.
 \label{odd:eq:uniform-time-variance}
\end{align}
The constants are uniform in $\delta,T,K$ and $N$.
\end{proposition}
\begin{proof}
We treat $f=1,2$ simultaneously. For prescribed nonempty words
$w_1,\ldots,w_f$ of total length $\ell$, write
$\mathcal D_{N,f}(w_1,\ldots,w_f)$ for the absolute sum of the
indicated normalized monomial expectations when $f=1$, and of their
covariances when $f=2$. In the second case, assignments with no shared
physical ordered entry have zero covariance and are omitted. An entry
of multiplicity one in the union also gives zero contribution, since
its entry variable occurs once and is independent of all others.
Thus every remaining assignment has connected physical support and
all entry multiplicities are at least two.

Fix such a canonical pattern. Let $e$ be its number of physical entries,
$v$ its number of physical vertices, and $s\ge1$ its number of odd
entry multiplicities. Put
\[
 d=\frac{\ell-s}{2}+1-v,\qquad g_{\rm ph}=e-v+1.
\]
The odd-entry subgraph is a nonempty even-degree multigraph, so
$g_{\rm ph}\ge1$ and $d\ge1$. If $\mu_m=\E\xi^m$, then
$|\mu_m|\le M^{m-2}$ for $m\ge2$. Hence the absolute joint moment
is at most $M^{\ell-2e}$.
For two faces, the product of the two separate moments is zero if an
entry occurs once in either face. Otherwise an entry occurring in both
faces, with multiplicities $a,b\ge2$, contributes at most
$M^{a-2}M^{b-2}\le M^{a+b-2}$; an entry occurring in just one face
satisfies the same joint-multiplicity bound. The absolute monomial
covariance is consequently at most $2M^{\ell-2e}$.

After summing the injective labels, the absolute normalized pattern
weight is bounded, in both cases, by
\begin{align*}
 2(N)_vN^{-f-\ell/2}M^{\ell-2e}
 &\le 2N^{1-f-d-s/2}M^{2d+s-2g_{\rm ph}}\\
 &\le 2N^{1-f-d-s/2}M^{2d+s}.
\end{align*}
This bound also applies in the one-face case.
Theorem~\ref{odd:thm:uniform-odd-count} therefore bounds
$\mathcal D_{N,f}$, for words of length $\ell\le2pfK$, by the finite
sum
\[
 \sum_{\substack{1\le s\le\lfloor\ell/3\rfloor\\
                  1\le d\le\lfloor(\ell-s)/2\rfloor}}
 C_{p,M}^{d+s}(1+\ell)^{pf+2d+s+f-3}
 N^{1-f-d-s/2}.
\]
Enlarging the sum to all the displayed values of $d$ and $s$ gives
an upper bound, regardless of parity or realizability. Put
$\alpha=C_{p,M}(1+\ell)^2/N$ and
$\beta=C_{p,M}(1+\ell)/\sqrt N$. Factoring out $d=s=1$
bounds the displayed sum by
\[
 C_{p,M}^{\,2}N^{-f-1/2}(1+\ell)^{pf+f}
 \sum_{j,k\ge0}\alpha^j\beta^k
 =\frac{C_{p,M}^{\,2}N^{-f-1/2}(1+\ell)^{pf+f}}
        {(1-\alpha)(1-\beta)}.
\]
Condition \eqref{odd:eq:uniform-time-cutoff} makes both ratios at
most $1/2$. Thus, uniformly on the cutoff,
\begin{equation}\label{odd:eq:uniform-word-bound}
 \mathcal D_{N,f}(w_1,\ldots,w_f)
 \le C_{p,M}N^{-f-1/2}(1+\ell)^{pf+f}.
\end{equation}
The empty word has no odd contribution, and an empty face gives a
deterministic monomial with zero covariance. Setting
$\mathcal D_{N,f}=0$ in these cases extends the bound to all words.
In a nonempty face we remove zero displayed blocks and retain the
labels of all other blocks, including consecutive blocks with the
same sign.

There are $pf$ time variables in the scalar expansion. Write
$u=\sum_{j=1}^{pf}t_j$. The two exponential blocks at time $t_j$
have lengths $a_j,b_j$ and coefficient
$t_j^{a_j+b_j}/(a_j!b_j!)$. Grouping the successive pairs of blocks
into $f$ words gives
\[
 R_f=\int_{[0,T]^{pf}}e^{-2(1+\delta)u}
 \sum_{\substack{0\le a_j,b_j\le K\\1\le j\le pf}}
 \left(\prod_{j=1}^{pf}\frac{t_j^{a_j+b_j}}{a_j!b_j!}\right)
 \mathcal D_{N,f}(w_1,\ldots,w_f)\dd t.
\]
All scalar coefficients are nonnegative. For fixed $t$, let
$A_j,B_j$ be independent Poisson variables of means $t_j$.
Their total $Z=\sum_j(A_j+B_j)$ is Poisson with mean $2u$.
After multiplying by $e^{-2u}$, the full coefficient sum is precisely
this law. Apply \eqref{odd:eq:uniform-word-bound} only on the
truncated set, and then extend its nonnegative scalar upper bound to
all block lengths. This yields
\[
 R_f\le C_{p,M}N^{-f-1/2}
 \int_{[0,T]^{pf}}e^{-2\delta u}\E(1+Z)^{pf+f}\dd t.
\]
The exponent $pf+f$ is fixed. Expansion into falling factorials and
$\E(Z)_j=(2u)^j$ give
$\E(1+Z)^{pf+f}\le C_p(1+u)^{pf+f}$. Since $u\le pfT$ and
$T\ge1$, the integral is at most $C_pT^{2pf+f}$, proving both
bounds. Summing the defects before extending the Poisson sum leaves
only the fixed-degree moment $\E(1+Z)^{pf+f}$ to estimate.
\end{proof}

In particular, fix $A\ge1$ and let $K=\lceil AT\rceil$.
If $T=o(\sqrt N)$ and $T\ge1$, then
\eqref{odd:eq:uniform-time-cutoff} holds eventually, and the critical
normalizations satisfy
\[
 T^{-2p+3/2}R_1\le C_{p,M}N^{-3/2}T^{5/2}\longrightarrow0,
 \qquad
 T^{-4p+3}R_2\le C_{p,M}N^{-5/2}T^5\longrightarrow0.
\]

\begin{theorem}[Growing-time comparison without symmetry]
\label{prop:general-growing-traces}
Assume \eqref{eq:entry-assumptions}. For every fixed integer $p\ge1$
there are positive constants $A_{p,M},B_{p,M},C_{p,M},c_{p,M}$ with
the following property. Let $\delta\ge0$, $T\ge1$, and
$K=\lceil A_{p,M}T\rceil$. If
\[
 B_{p,M}(K+1)^2/N\le\tfrac12,\qquad T\le c_{p,M}N,
\]
then, with $\zeta_{p,M}=C_{p,M}(e^{-c_{p,M}T}+e^{-c_{p,M}N})$,
\begin{align}
 \left|\E\tau_NQ_{\delta,T}(X_N)^p-\tau(q_{\delta,T}^p)\right|
 &\le C_{p,M}\left(N^{-1}T^{2p+1/2}
                  +N^{-3/2}T^{2p+1}\right)+\zeta_{p,M},
 \label{eq:general-growing-mean}\\
 \operatorname{Var}\bigl(\tau_NQ_{\delta,T}(X_N)^p\bigr)
 &\le C_{p,M}\left(N^{-2}T^{4p+1/2}
                  +N^{-5/2}T^{4p+2}\right)+\zeta_{p,M}.
 \label{eq:general-growing-var}
\end{align}
\end{theorem}
\begin{proof}
Use the polynomial Gramian $A_N$ in Proposition~\ref{prop:odd-time-sums},
and its circular counterpart $A_c$. Let $\mathcal E_{p,N}$ be the
contribution of the even physical-entry patterns to $\E\tau_N A_N^p$.
Lemmas~\ref{lem:bounded-even} and \ref{lem:bounded-planar} apply to these
patterns for every entry law satisfying \eqref{eq:entry-assumptions}.
The finite deficit sum and Poisson coefficient calculation in the proof
of Theorem~\ref{prop:bounded-traces} therefore give
\[
 |\mathcal E_{p,N}-\tau(A_c^p)|
       \le C_{p,M}N^{-1}T^{2p+1/2}.
\]
In that calculation, symmetry is used only to identify the even part
with the whole expectation. Applying the same deficit and Poisson
summations to \eqref{eq:bounded-even-var} bounds the sum of the
absolute monomial covariances with even union multiplicities by
$C_{p,M}N^{-2}T^{4p+1/2}$. This includes entries whose
multiplicities in the individual faces are both odd: the comparison
bounds the joint moment and the product of the separate moments
together.

The remaining terms are bounded in absolute value by $R_1$ and $R_2$
from Proposition~\ref{prop:odd-time-sums}. Increasing $B_{p,M}$ makes
both cutoff inequalities in \eqref{odd:eq:uniform-time-cutoff} hold.
Adding its bounds proves both displayed
inequalities for the polynomial Gramians.

The polynomial approximation in Theorem~\ref{prop:bounded-traces}
uses only boundedness, centering and independence of the entries.
Its norm tail and exponential remainder therefore apply here unchanged.
They bound the $L^1$ approximation error and the squared $L^2$ error
by $\zeta_{p,M}$. The circular remainder has the same bound.
The mean comparison follows by the triangle inequality, and
$\operatorname{Var}Z\le2\operatorname{Var}W+2\E|Z-W|^2$
transfers the variance estimate after adjusting the constants.
\end{proof}

\begin{corollary}\label{cor:general-fixed-L}
Assume \eqref{eq:entry-assumptions}. If $\delta_N\to0$ is deterministic
and $\sqrt N\delta_N\to\infty$, then for every fixed integer $p\ge1$
and every $L>0$,
\[
 \delta_N^{2p-3/2}\tau_NQ_{N,\delta_N,L/\delta_N}^{p}
       \xrightarrow{\Prob}D_p(L).
\]
\end{corollary}
\begin{proof}
Set $T=L/\delta_N$ and $a_N=\sqrt N\delta_N$. The conditions of
Theorem~\ref{prop:general-growing-traces} hold eventually. After
multiplication by $\delta_N^{2p-3/2}$, the two mean errors are bounded
by constants depending on $p,M,L$ times
\[
 a_N^{-2},\qquad N^{-1/4}a_N^{-5/2}.
\]
After multiplication by $\delta_N^{4p-3}$, the two variance terms
are bounded by constants times
\[
 N^{-1/4}a_N^{-7/2},\qquad a_N^{-5}.
\]
All four quantities tend to zero. The exponential remainders also
vanish after normalization, since $T\to\infty$ and $T=o(\sqrt N)$.
The circular finite-time limit in Lemma~\ref{lem:finite-free} identifies
the limiting mean as $D_p(L)$. Chebyshev's inequality completes the proof.
\end{proof}

\section{Energy from Hermitization}
\label{sec:real-energy}

The trace estimates so far concern times of order $\delta_N^{-1}$.
To reach the stationary covariance, we need a bound on the energy
remaining after such a time. Plancherel's identity expresses the total
energy through the resolvent on a vertical line outside the spectrum.
Exterior local laws for the singular values of $X_N-zI$ then give the
required bound with a fixed constant. Throughout this section, $z$
denotes the matrix shift and $i\eta$ the Hermitian spectral parameter.

\begin{proposition}\label{prop:real-energy}
Fix $0<\varepsilon<1/2$ and let $\delta_N\to0$ satisfy
$\delta_N\ge N^{-1/2+\varepsilon}$. Assume the bounded-entry hypotheses
in \eqref{eq:entry-assumptions}.
There is a constant $C=C_M$ such
that, for every $D>0$, there are events $\mathcal A_N$ with
$\Prob(\mathcal A_N^c)\le C_{\varepsilon,D}N^{-D}$ for all
sufficiently large $N$, on which $X_N$ is stable at damping
$1+\delta_N/2$ and
\begin{equation}\label{eq:energy-hp}
 \delta_N^{1/2}\tau_NQ_{N,\delta_N/2}\le C.
\end{equation}
For every $L>0$, on the same event,
\begin{equation}\label{eq:positive-tail}
 0\le Q_{N,\delta_N}-Q_{N,\delta_N,L/\delta_N}
       \le e^{-L}Q_{N,\delta_N/2}.
\end{equation}
In particular, its trace on the critical energy scale satisfies
\begin{equation}\label{eq:scaled-time-tail}
 \delta_N^{1/2}\tau_N
 [Q_{N,\delta_N}-Q_{N,\delta_N,L/\delta_N}]
 \le Ce^{-L}.
\end{equation}
\end{proposition}

\begin{proof}
Write
\[
 \begin{gathered}
 H^z=\begin{pmatrix}0&X_N-zI\\X_N^*-\overline z I&0\end{pmatrix},
 \qquad G^z(i\eta)=(H^z-i\eta I)^{-1},\\
 m_N(z,\eta)=\frac1{2N}\tr G^z(i\eta).
 \end{gathered}
\]
For the singular values $s_j(z)$ of $X_N-zI$,
\begin{equation}\label{eq:real-hermitized-trace}
 \Im m_N(z,\eta)
   =\frac{\eta}{N}\sum_{j=1}^N\frac1{s_j(z)^2+\eta^2}.
\end{equation}
Here $m_N$ is normalized in dimension $2N$, whereas $\tau_N$ is
normalized in dimension $N$. For the constant variance matrix
$S_{ij}=1/N$, uniqueness and coordinate symmetry in the vector
Dyson equation of \cite[Eq.~(2.3)]{AEK21} identify both of its
components with a scalar $v>0$. The deterministic approximation
to $m_N$ is $iv$, where
\[
 \frac1v=\eta+v+\frac{|z|^2}{\eta+v}.
\]
Putting $\kappa=|z|^2-1>0$, this equation gives the exact identity
and bound
\begin{equation}\label{eq:real-dyson-upper}
 \eta=v\bigl[\kappa+(\eta+v)^2\bigr],
 \qquad v\le\frac{\eta}{\kappa}.
\end{equation}

We first work near the unit circle. The exterior averaged local law
of \cite[Theorem~5.2, Eq.~(5.4)]{AEK21} states that
\begin{equation}\label{eq:real-exterior-local-law}
 |m_N(z,\eta)-iv|
   \prec \frac{N^{-\gamma/3}}{N\eta}
\end{equation}
uniformly when
\[
 |z|\le1+\tau_*,\qquad
 \kappa\ge(N^\gamma\eta)^{2/3},\qquad
 N^\zeta\eta_f(z)\le\eta\le\tau_*,
 \qquad
 \eta_f(z)=N^{-2/3}\kappa^{1/6}.
\]
Here $\tau_*>0$ is fixed and may be decreased so that
$(1+\tau_*)^2\le2$; $\gamma,\zeta>0$ are fixed, and
$A_N\prec B_N$ means that
$\Prob(A_N>N^\xi B_N)\le C_{\xi,D}N^{-D}$ for every
$\xi,D>0$. The formula for $\eta_f$ applies in the exterior
regime $\kappa>N^{-1/2}$ under consideration. The law holds for real
entries and every argument of $z$ under assumptions (A1) and (A2)
of that paper: flat variances and bounded moments of every fixed
normalized entry order. Both follow from
\eqref{eq:entry-assumptions}. The scalar equation above uses the
simplified self-energy convention of \cite{AEK21}; its Remark~5.4
absorbs the additional terms from $\E (X_N)_{ij}^2=1/N$ into the
error estimate. Thus $iv$ is the deterministic approximation for
the real iid model as well.

Let $\theta=\varepsilon/2$ and suppose
$\kappa\ge N^{-1/2+\theta}$. Choose
\[
 \eta=N^{-\theta/12}\sqrt{\frac{\kappa}{N}},
 \qquad \gamma=\theta,\qquad\zeta=\theta/12.
\]
The required inequalities follow from
\[
 \frac{\eta}{\eta_f}
   =N^{-\theta/12}(N\kappa^2)^{1/6}\ge N^{\theta/4},
 \qquad
 \frac{N^\theta\eta}{\kappa^{3/2}}
   =\frac{N^{11\theta/12}}{\sqrt N\,\kappa}\le N^{-\theta/12}.
\]
Also $\eta\le\tau_*$ for large $N$. Fix $0<\xi<\theta/12$.
Outside an event of probability $C_{\varepsilon,D}N^{-D}$,
\eqref{eq:real-exterior-local-law} and
\eqref{eq:real-dyson-upper} therefore give
\[
 N\eta\,\Im m_N
 \le \frac{N\eta^2}{\kappa}+N^{\xi-\theta/3}
 =N^{-\theta/6}+N^{\xi-\theta/3}=o(1).
\]
If some $s_j(z)\le\eta$, the left side would be at least
$1/2$ by \eqref{eq:real-hermitized-trace}. Thus every
$s_j(z)>\eta$. Consequently, for $R(z)=(zI-X_N)^{-1}$,
\begin{align}
 \tau_NR(z)R(z)^*
 &\le \frac{2\Im m_N(z,\eta)}{\eta}\notag\\
 &\le \frac2\kappa
      +\frac{2N^{\xi-\theta/3}}{N\eta^2}
 =\frac2\kappa\bigl(1+N^{\xi-\theta/6}\bigr)
 \le\frac3\kappa. \label{eq:real-near-edge-energy}
\end{align}
The choice $\xi<\theta/12$ makes the final error tend to zero,
which yields the fixed constant in this bound.

We also need a bound on a fixed exterior annulus
$1+c\le|z|\le B$, where $c>0$ and $B$ are fixed.
Here \cite[Theorem~5.2, Eq.~(5.4)]{AEK18} gives
\[
 |m_N(z,\eta)-iv|\prec
       \frac1N+\frac1{(N\eta)^2}.
\]
The theorem permits nonzero $\E(X_N)_{ij}^2$ and therefore applies
to real matrices under the same variance and moment assumptions.
Choose $\eta=N^{-7/12}$. Taking its domain exponent to be $1/4$
gives $N^{-1+1/4}\le\eta\le1$, as required.
Fix $0<\xi<1/8$. Since $\kappa$ is
bounded above and below on the annulus,
\[
 N\eta\,\Im m_N
 \le C N^{-1/6}
       +N^\xi\bigl(N^{-7/12}+N^{-5/12}\bigr)=o(1),
\]
while
\[
 \frac{\Im m_N}{\eta}
 \le C+N^\xi\bigl(N^{-5/12}+N^{-1/4}\bigr)\le C+1.
\]
The preceding singular-value argument gives
$s_{\min}(X_N-zI)>\eta$ and
$\tau_NR(z)R(z)^*\le C$ on this annulus, with very high
probability.

We next make the resolvent bounds simultaneous along the integration
line. Put $a=1+\delta_N/2$ and
$z(\omega)=a+i\omega$. Then
\[
 \kappa(\omega)=a^2-1+\omega^2
   \ge\delta_N+\omega^2\ge N^{-1/2+\theta}
\]
for large $N$. Set $K_M=10M$. Cover $|\omega|\le2K_M$ by a deterministic grid
of mesh $N^{-3}$. Apply the preceding estimates at every grid
point, using the near-edge region through radius $1+\tau_*$
and the fixed annulus with $c=\tau_*/4$ and $B=2K_M+2$.
These cover the grid for large $N$ and overlap around the dividing
radius $1+\tau_*/2$.
A union bound with the failure exponent increased by four
gives one event of probability $1-O_D(N^{-D})$ on which all
grid estimates hold. In the near-edge region the chosen regularization
satisfies $\eta\ge N^{-3/4+5\theta/12}>N^{-1}$; on the fixed annulus
it is $N^{-7/12}$. Thus every grid resolvent has norm at most $N$.
If $|z-z_j|\le N^{-3}$, the identity
\[
 R(z)=\bigl[I+(z-z_j)R(z_j)\bigr]^{-1}R(z_j)
\]
shows that $zI-X_N$ is invertible and
$\norm{R(z)}_{\mathrm{HS}}\le2\norm{R(z_j)}_{\mathrm{HS}}$.
Furthermore $\kappa(z)/\kappa(z_j)=1+o(1)$ uniformly.
We have therefore obtained, on a single event,
\[
 \tau_NR(a+i\omega)R(a+i\omega)^*
 \le
 \begin{cases}
 C/(\delta_N+\omega^2),&|a+i\omega|\le1+\tau_*/2,\\
 C,&|\omega|\le2K_M,\ |a+i\omega|\ge1+\tau_*/2.
 \end{cases}
\]
The overlap in the two local-law domains justifies the indicated
choice of dividing radius.

For real unit vectors $u,v$, Hoeffding's inequality and independence give
\[
 \E e^{t u^{\mathsf T}X_Nv}\le e^{M^2t^2/(2N)},\qquad
 \Prob(|u^{\mathsf T}X_Nv|>s)\le2e^{-Ns^2/(2M^2)}.
\]
Two $1/8$-nets of the real unit sphere have size at most $17^N$
each. The net approximation to the operator norm therefore gives
\[
 \Prob(\|X_N\|>x)
 \le2\,17^{2N}e^{-9Nx^2/(32M^2)}\le e^{-Nx^2/(8M^2)},\qquad x\ge K_M.
\]
On $\|X_N\|\le K_M$, of probability at least $1-e^{-cN}$,
the Neumann series yields
\[
 \norm{R(a+i\omega)}
 \le\frac1{|a+i\omega|-K_M}\le\frac2{|\omega|}
 \qquad (|\omega|\ge2K_M).
\]
This controls the remaining frequency integral by a fixed constant.

To apply Plancherel, we also use the spectral-radius bound of
\cite[Remark~2.2]{AEK21}. Under the same variance and moment
conditions it gives, with very high probability,
\[
 \max_{\lambda\in\operatorname{spec}(X_N)}|\lambda|
 \le1+N^{-1/2+\varepsilon/2}<1+\delta_N/2=a.
\]
Intersect this event with the preceding events to define
$\mathcal A_N$. On $\mathcal A_N$, each entry of the stable finite-dimensional
semigroup is in $L^1\cap L^2$ on positive times, as follows by
writing its Jordan blocks as polynomials multiplying exponentially
decaying functions. Its Fourier transform is $R(a+i\omega)$.
Entrywise Plancherel therefore gives
\[
 \tau_NQ_{N,\delta_N/2}
   =\frac1{2\pi}\int_{\mathbb R}
       \tau_NR(a+i\omega)R(a+i\omega)^*\,\dd\omega
   \le C\int_{\mathbb R}\frac{\dd\omega}{\delta_N+\omega^2}+C
   \le C'\delta_N^{-1/2}.
\]
This proves \eqref{eq:energy-hp}.
Finally, for $t\ge L/\delta_N$,
\[
 e^{-2(1+\delta_N)t}
  =e^{-\delta_Nt}e^{-2(1+\delta_N/2)t}
  \le e^{-L}e^{-2(1+\delta_N/2)t}.
\]
Integrating the positive matrix integrand proves
\eqref{eq:positive-tail} and its trace bound.
\end{proof}

The event $\mathcal A_N$ is independent of $L$. We can therefore
first take the matrix limit at fixed rescaled time and then let that
time tend to infinity, with an error bounded by $Ce^{-L}$ throughout.

\section{Passing to the stationary spectral law}
\label{sec:transfer}

We now combine the finite-time moment limits with the energy bound.
At each fixed rescaled time, the moments determine a compactly
supported limit. The positive time tail then controls the difference
between this measure and the stationary one. Once the weighted law
has been obtained, fixed spectral thresholds recover both the ranks
and their cumulative energy.

\begin{lemma}[Convergence from moments]\label{lem:random-moments}
Let $\xi_n$ be random positive finite measures on $[0,\infty)$,
and let $\xi$ be a deterministic compactly supported finite measure.
If every nonnegative integer moment of $\xi_n$ converges in
probability to the corresponding moment of $\xi$, then
$\xi_n\Rightarrow\xi$ in probability.
\end{lemma}
\begin{proof}
From any subsequence, extract a further subsequence on which all
countably many moments converge almost surely. For every outcome
in this probability-one event, bounded zeroth and first moments
give tightness of the deterministic sequence of measures. A bound
on the next moment makes each fixed moment uniformly integrable:
\[
 \int_{x>R}x^k\,\xi_n(\dd x)
 \le R^{-1}\int x^{k+1}\,\xi_n(\dd x).
\]
Consequently every weak subsequential limit has the moments of
$\xi$. If $\xi$ is supported on $[0,M]$, these moments force any
measure $\eta$ with the same moments to be supported there:
\[
 \eta((M+\varepsilon,\infty))
 \le \xi([0,\infty))
       \left(\frac M{M+\varepsilon}\right)^k\longrightarrow0.
\]
The degenerate case $M=0$ follows from the first moment.
Uniform polynomial approximation on the compact interval then
gives uniqueness. Thus the extracted sequence converges weakly
almost surely. The subsequence characterization proves convergence
in probability of the original sequence.
\end{proof}

\begin{proof}[Proof of Theorem~\ref{thm:matrix-crossover}]
Write $\delta=\delta_N$, and let $\Omega_N$ be the event in
Proposition~\ref{prop:real-energy}, whose hypotheses follow from
\eqref{eq:window}.
For each fixed $L>0$, define the finite-time weighted measure by
\[
 \int f\,\dd\nu_{N,L}
   =\delta^{1/2}\tau_N[
       Q_{N,\delta,L/\delta}f(\delta^2Q_{N,\delta,L/\delta})].
\]
It exists on every sample, even if the stationary matrix does not.
Since $\delta_N\sqrt N\to\infty$,
Corollary~\ref{cor:general-fixed-L} applies. Thus, for each fixed
$k\ge0$,
\[
 \int x^k\,\nu_{N,L}(\dd x)
 =\delta^{2k+1/2}\tau_NQ_{N,\delta,L/\delta}^{k+1}
 \xrightarrow{\Prob}D_{k+1}(L).
\]
Lemma~\ref{lem:finite-free} identifies these as the moments of the
compactly supported $\nu_L$. By Lemma~\ref{lem:random-moments},
\begin{equation}\label{eq:finite-cut-weak}
 \nu_{N,L}\Rightarrow\nu_L
 \quad\hbox{in probability for every fixed }L>0.
\end{equation}

To remove the time cutoff, observe that for Hermitian $A\ge B$ and a scalar Lipschitz
function $g$, eigenvalue monotonicity gives
\begin{equation}\label{eq:trace-lipschitz}
 |\tau_Ng(A)-\tau_Ng(B)|
 \le \operatorname{Lip}(g)\tau_N(A-B).
\end{equation}
Indeed the ordered eigenvalues of $A$ dominate those of $B$,
and summing the scalar Lipschitz inequalities proves the formula.

If $f$ is a compactly supported Lipschitz function on $[0,\infty)$,
then $g(x)=xf(x)$ is globally Lipschitz, and the weighting gives
\[
 \nu_{N,\delta}(f)
   =\delta^{-3/2}\tau_N g(\delta^2Q_{N,\delta}),
 \qquad
 \nu_{N,L}(f)
   =\delta^{-3/2}\tau_N g(\delta^2Q_{N,\delta,L/\delta}).
\]
Apply \eqref{eq:trace-lipschitz} with
$A=\delta^2Q_{N,\delta}$ and
$B=\delta^2Q_{N,\delta,L/\delta}$ on $\Omega_N$. The positive time-tail bound of
Proposition~\ref{prop:real-energy} gives, on $\Omega_N$,
\begin{equation}\label{eq:test-tail}
 |\nu_{N,\delta}(f)-\nu_{N,L}(f)|
 \le C\operatorname{Lip}(g)e^{-L}.
\end{equation}
For every fixed $D>0$ the complement has probability
$O_{\varepsilon,D}(N^{-D})$. Explicitly, for any tolerance $a>0$,
\[
 \Prob\bigl(|\nu_{N,\delta}(f)-\nu_{N,L}(f)|>a\bigr)
 \le \Prob(\Omega_N^c)
       +\one_{\{C\operatorname{Lip}(g)e^{-L}>a\}}.
\]
For $f=1$ the same bound holds with $g(x)=x$ and Lipschitz
constant one.

Compare $\nu_{N,\delta}(f)$ successively with $\nu_{N,L}(f)$,
$\nu_L(f)$ and $\nu_*(f)$. At fixed $L$ the middle comparison tends
to zero in probability by \eqref{eq:finite-cut-weak}. The last
comparison tends to zero as $L\to\infty$ by
Lemma~\ref{lem:finite-free}. The first is controlled by
\eqref{eq:test-tail}. This proves convergence against every
compactly supported Lipschitz $f$, and separately proves convergence
of total masses. In particular it proves \eqref{eq:first-mass}.

The compact tests and the total mass together give weak convergence
in probability.
Choose a compactly supported Lipschitz cutoff $0\le\chi\le1$ that
equals one on $[0,M]$ and vanishes on $[M+1,\infty)$, where $M>1/2$.
Since $\nu_*$ is supported on $[0,1/2]$,
\[
 \nu_{N,\delta}((M+1,\infty))
 \le \nu_{N,\delta}(1)-\nu_{N,\delta}(\chi)
 \xrightarrow{\Prob}0 .
\]
Together with boundedness of the masses in probability, this gives
tightness. More explicitly, from every subsequence we can extract a
further subsequence on which the total mass and a countable
determining family of compact tests converge almost surely. On that
subsequence, deterministic tightness and uniqueness give weak
convergence. The subsequence criterion then proves the assertion in
probability.

For $x>0$ the exact identity
\begin{equation}\label{eq:tail-identity-N}
 H_N(x):=
 \frac{\#\{j:\lambda_j(Q_{N,\delta})>x\delta^{-2}\}}
      {N\delta^{3/2}}
 =\int_{(x,\infty)}y^{-1}\nu_{N,\delta}(\dd y)
\end{equation}
holds also on samples where the stationary matrix was set to zero.
The integrand is bounded and has just one discontinuity, at $x$.
The limiting beta-product measure has no atom there. Applying weak
convergence along almost-sure subsequences proves
$H_N(x)\to H(x)$ in probability for every $x>0$.

On any $[a,b]\subset(0,\infty)$, choose a finite partition so that
the oscillation of the continuous $H$ on each interval is less
than $\varepsilon$. With probability tending to one, the values of
$H_N$ at all partition endpoints are within $\varepsilon$ of $H$.
Monotonicity of $H_N$ then bounds the sup norm difference by
$2\varepsilon$. This proves \eqref{eq:counts}, including intervals
crossing the endpoint $1/2$.

The beta-product density is positive on $(0,1/2)$, so $H$ is
strictly decreasing there. Also $\E U^{-1}=\infty$, and independence
implies $H(x)\uparrow\infty$ as $x\downarrow0$ by monotone
convergence. Thus $x_r=H^{-1}(r)$ is defined for every $r>0$.
Under the window, $N\delta_N^{3/2}\to\infty$ and
$N\delta_N^{3/2}=o(N)$, so the stated ranks are eventually valid. For fixed points $x_-<x_r<x_+$
inside $(0,1/2)$, one has $H(x_-)>r>H(x_+)$.
The counts at these two points bracket $k_N$ with probability
tending to one. Hence $x_-<\delta_N^2\lambda_{k_N}\le x_+$
with that probability. Letting the two points approach $x_r$
proves \eqref{eq:quantile}.
\end{proof}

\begin{proof}[Proof of Corollary~\ref{cor:explained-variance}]
Write $x_j=\delta_N^2\lambda_j(Q_{N,\delta_N})$ and
$\nu_N=\nu_{N,\delta_N}$. Then
\begin{equation}\label{eq:partial-energy-normalization}
 \nu_N=\frac1{m_N}\sum_jx_j\delta_{x_j},\qquad
 S_N(k):=\frac1{m_N}\sum_{j=1}^kx_j
       =\frac{\delta_N^{1/2}}{N}\sum_{j=1}^k\lambda_j(Q_{N,\delta_N}).
\end{equation}
On stable samples $Q_{N,\delta_N}$ is positive definite: its
integrand is continuous and equals $I$ at time zero. Thus $F_N(k)$
and $K_N(\theta)$ are well defined, with
\[
 F_N(k)=\frac{S_N(k)}{\nu_N([0,\infty))}.
\]

To control a partial sum with a moving rank, we bracket it by energy
tails at fixed thresholds. Fix $r>0$ and set $x_r=H^{-1}(r)$.
Choose
$0<x_-<x_r<x_+<1/2$. Since $H(x_-)>r>H(x_+)$, the pointwise
count convergence proved above and $k_N/m_N\to r$
give, with probability tending to one,
\[
 \#\{j:x_j>x_+\}<k_N<\#\{j:x_j>x_-\}.
\]
Every value above $x_+$ belongs to the first $k_N$ entries, and
all of these entries exceed $x_-$. Thus, including when eigenvalues
are repeated,
\begin{equation}\label{eq:partial-energy-sandwich}
 \nu_N((x_+,\infty))\le S_N(k_N)\le\nu_N((x_-,\infty)).
\end{equation}
The two tail indicators are bounded and their discontinuities
have zero $\nu_*$-mass. Weak convergence therefore gives
\[
 \nu_N((x_\pm,\infty))\xrightarrow{\Prob}
                    \nu_*((x_\pm,\infty)).
\]
For any given tolerance, choose $x_-$ and $x_+$ close enough to
$x_r$ that both limiting tails differ from
$\nu_*((x_r,\infty))$ by less than half that tolerance.
Taking $N\to\infty$ with these thresholds fixed in
\eqref{eq:partial-energy-sandwich} proves
\[
 S_N(k_N)\xrightarrow{\Prob}\nu_*((x_r,\infty)).
\]
The total mass converges in probability to $C_1>0$, and hence
exceeds $C_1/2$ with probability tending to one. Dividing the
partial-sum limit by the total-mass limit proves
\eqref{eq:explained-variance}. The convention on unstable samples
does not affect this limit, since their probability tends to zero.
The positive density of $Y$ on $(0,1/2)$, together with
$H^{-1}(0+)=1/2$ and $H^{-1}(\infty)=0$, gives the stated
properties of $\mathcal E$.

For a prescribed variance fraction $0<\theta<1$, the same argument
can be inverted using deterministic ranks. Choose constants
$0<r_-<r_\theta<r_+$. Set
$k_N^\pm=\lfloor r_\pm m_N\rfloor$. These are valid ranks for
all sufficiently large $N$, since $m_N\to\infty$ and $m_N=o(N)$.
Applying \eqref{eq:explained-variance} at these two deterministic
rank sequences, and using
$\mathcal E(r_-)<\theta<\mathcal E(r_+)$, gives
\[
 F_N(k_N^-)<\theta<F_N(k_N^+)
\]
with probability tending to one. On this event,
$k_N^-<K_N(\theta)\le k_N^+$. Divide by $m_N$ and let the two
fixed constants approach $r_\theta$ to obtain
\eqref{eq:retained-dimension}.

For the reconstruction interpretation, fix a stable matrix
realization and let $Z$ have its centered stationary distribution.
For an orthogonal projection $P=P(X_N)$ chosen from that realization,
\[
 \E[\|Z-PZ\|^2\mid X_N]=\tr Q-\tr(PQ).
\]
The maximum of $\tr(PQ)$ over rank-$k$ orthogonal projections is
the sum of the $k$ largest eigenvalues of $Q$. Consequently,
\[
 \min_{\substack{P=P^{\mathsf T}=P^2\\\operatorname{rank}P=k}}
 \frac{\E[\|Z-PZ\|^2\mid X_N]}{\E[\|Z\|^2\mid X_N]}
 =1-F_N(k).
\]
Equation~\eqref{eq:explained-variance} gives the asserted error limit.

To obtain \eqref{eq:compression-tail}, let $\rho$ be the density of
$\nu_*$. The endpoint calculation in the proof of
Corollary~\ref{free:cor:ends} gives
$\rho(x)\sim(3c_0/4)x^{-3/4}$ as $x\downarrow0$. Since
$C_1=1/(2\sqrt2)$, integration yields
\[
 \Prob(Y\le x)=C_1^{-1}\int_0^x\rho(y)\dd y
       \sim6\sqrt2\,c_0x^{1/4}.
\]
Substitute $x=H^{-1}(r)\sim(c_0/r)^{4/3}$ and use
$1-\mathcal E(r)=\Prob\{Y\le H^{-1}(r)\}$ to obtain the stated
asymptotic for the limiting compression curve.
\end{proof}

\paragraph*{Control energy}
Fix $N$, $\delta$ and a stable realization, and put
$A=X_N-(1+\delta)I$. For $T>0$, the endpoint map from
$L^2([0,T];\R^N)$ to $\R^N$ is
\[
 \mathcal R_Tu=\int_0^Te^{A(T-s)}u(s)\dd s.
\]
Its adjoint is $(\mathcal R_T^*v)(s)=e^{A^{\mathsf T}(T-s)}v$,
so $\mathcal R_T\mathcal R_T^*=Q_{N,\delta,T}$. This matrix is
positive definite: for $v\ne0$,
\[
 v^{\mathsf T}Q_{N,\delta,T}v
   =\int_0^T\|e^{A^{\mathsf T}s}v\|^2\dd s>0.
\]
The control $u_* =\mathcal R_T^*Q_{N,\delta,T}^{-1}v$ reaches
$v$ from the origin. Every other such control differs from $u_*$
by an element of $\ker\mathcal R_T$, which is orthogonal to
$u_*$. Pythagoras therefore yields
\begin{equation}\label{eq:control-energy}
 \min_{\substack{\dot z=Az+u\\ z(0)=0,\ z(T)=v}}
       \int_0^T\|u(s)\|^2\dd s
       =v^{\mathsf T}Q_{N,\delta,T}^{-1}v.
\end{equation}
At this fixed realization, stability gives
$Q_{N,\delta,T}\to Q_{N,\delta}>0$ in norm as $T\to\infty$.
Continuity of matrix inversion gives the limiting cost
$v^{\mathsf T}Q_{N,\delta}^{-1}v$, which equals $1/\lambda$
when $v$ is a unit eigenvector of $Q_{N,\delta}$ with eigenvalue
$\lambda$.

\paragraph*{Variance concentration near instability}
In the simultaneous limit, the number of leading principal components
needed to retain any fixed fraction of the stationary variance diverges,
while its ratio to $N$ tends to zero. The explicit profile describes
both this concentration and the variance left after projection.
Along the limiting curve, the $r^{-4/3}$ eigenvalue tail gives an
$r^{-1/3}$ decay of the residual variance. The covariance spectrum
therefore quantifies the dimension needed to represent the accumulated
response to noise. The product relation with the discrete-time law
accounts for the different rank exponents in continuous and discrete time.

\begin{appendix}

\section{Gaussian traces on growing time intervals}\label{sec:gaussian}

Let $G_N$ be a real Ginibre matrix with entry variance $1/N$, and
write $Q^G_{N,\delta,T}=Q_{\delta,T}(G_N)$. The comparison argument
in Section~\ref{sec:comparison} requires estimates on time intervals
that grow with $N$. We obtain them by counting Wick pairings at fixed
block lengths and then summing the exponential coefficients.

We use $+$ for an occurrence of $G_N$ and $-$ for one of
$G_N^{\mathsf T}$. A real Wick pairing identifies the physical row
indices of its two occurrences, and likewise their column indices.
In the polygon representation this reverses the boundary order for
opposite signs and preserves it for equal signs. The resulting closed
surface may be nonorientable. For a connected map with $V$ index
vertices, $n$ paired edges and $f$ trace faces, put
\[
 h=2-(V-n+f)\ge0.
\]
Thus $h=2g$ in the orientable case and $h$ is the number of crosscaps
otherwise. The exact index-sum weight, including normalized traces,
is
\begin{equation}\label{gauss:eq:real-weight}
 N^{V-n-f}=N^{2-2f-h}.
\end{equation}
For one face, $h=0$ means a sphere. Its polygon boundary orientation
makes every gluing reverse the boundary order. Every pair therefore
has opposite signs, and these are exactly the circular noncrossing
pairings. The leading one-face contribution is therefore the circular moment.

\begin{lemma}[Prescribed block lengths]\label{gauss:lem:real-count}
Fix $R\ge2$. Consider $f\in\{1,2\}$ prescribed rooted nonempty
cyclic sign words with $2n$ letters and $r\le R$ cyclic sign changes
in total. There is a constant $K_R$ with the following bounds for
connected real Wick pairings of deficit $h$.
Form the block graph by giving each nonempty maximal constant-sign
block a vertex, and joining the blocks of the two occurrences of each
edge. A pure face supplies one block, and loops are allowed. Put
$D=2h+2f-3+r/2$. Then
\begin{align}
 \#\{\text{pairings with a bipartite block graph}\}
      &\le K_R^{h+1}(n+1)^D,\label{gauss:eq:count-bip}\\
 \#\{\text{pairings with a nonbipartite block graph}\}
      &\le K_R^{h+1}(n+1)^{D-1}.\label{gauss:eq:count-nonbip}
\end{align}
Every pairing in the first class imposes a signed balance
$\sum_j\eta_j a_j=0$ on the block lengths, with $\eta_j\in\{1,-1\}$.
The bounds include pure faces and zero displayed blocks, after the
latter are deleted and adjacent blocks of the same sign are merged.
\end{lemma}

\begin{proof}
\emph{Compression and reconstruction.}
Rotate each mixed face to a prescribed sign-change corner, with the
rotation determined by the word. Keep the root of each pure face,
and let $t$ be the number of pure faces. Mark all vertices containing
a sign-change corner or a root. The following endpoint description
works on both orientable and nonorientable surfaces. Give every boundary position $j$ two endpoint
flags $(j,0),(j,1)$, in boundary order, and let $\varphi$ follow the
face lists. The corner involution joins $(j,1)$ to $(\varphi j,0)$.
The pairing involution on flags joins the two physical row endpoints
of each paired occurrence, and separately its two column endpoints.
For a positive occurrence the row flag is $(j,0)$; for a negative
occurrence it is $(j,1)$. These two involutions form alternating
cycles, which are exactly the index-vertex links.

Along a pairing link the row/column type is preserved. A corner
link changes that type precisely when its two letters have equal
signs. Hence a vertex of degree $d_v$ with $s_v$ sign-change corners
satisfies $d_v-s_v\equiv0\pmod2$. An unmarked vertex therefore has
even degree. At an unmarked degree-two vertex, its two physical
edge ends have opposite types: one edge enters and one leaves in
the row-to-column direction. This gives the subdivision a consistent
direction on either type of surface.

Retain marked vertices and unmarked vertices of degree at least four.
Suppress maximal chains of unmarked degree-two vertices and record
their positive lengths. This reverses edge subdivision and preserves
connectedness, faces, and $h$. At both occurrences of a suppressed
vertex the adjacent letters belong to the same original block.
Thus an entire chain has its two sides in fixed blocks, and
suppression preserves those sides. An unmarked degree-two vertex
incident twice
to the same loop would form an isolated component. Connectedness
would then make it the whole graph, contradicting the presence of a
marked face root. Every permitted suppression therefore joins two
distinct edges and removes one vertex and one edge.
A graph consisting entirely of a cycle retains all its face roots;
its core is a rooted cycle, possibly one loop. With two faces, two
distinct root vertices can remain. In particular, a nonempty pairing
never has an empty core. Twisted loops are allowed.

To recover the pairing, retain the ordered rooted core-face lists
and the sign of each core occurrence. Replace an edge of length
$l$ by two runs of $l$ letters. If the signs of its occurrences are
opposite, pair position $j$ with position $l+1-j$ in the other run;
if they agree, pair position $j$ with position $j$. These are the
unique expansions into a directed chain with unmarked degree-two
interior vertices. The two core occurrences remain distinct even
when they are consecutive occurrences of the same loop in one pure
face: their separating retained corner belongs to the rooted face
list. A root in either face cuts a chain, so its position cannot drift
inside a reconstructed run. The roots fix the first positions in the
face lists, and undoing the prescribed rotations recovers the original
labelled pairing. For a two-face cycle with distinct roots, the two
chain lengths determine the relative displacement of the roots. The
expansion also applies to a loop, or to two occurrences in the same
face. The signs fix every gluing along the chain, giving a unique expansion.

\emph{The cost of connecting block components.}
Let $e$ be the number of core edges. At a core vertex $v$, let $d_v$
be its degree and $s_v$ the number of sign-change corners. Write
$S=\{v:s_v>0\}$ and let $\mathcal R$ be the root-only vertices.
Then $|\mathcal R|\le t$, vertices in $\mathcal R$ have degree at
least two, and every other retained vertex outside $S$ has degree
at least four. If there are $u$ such other vertices, Euler's formula
and the degree sum read
\[
 e=u+|S|+|\mathcal R|+h+f-2,\qquad
 2e\ge4u+\sum_{v\in S}d_v+2|\mathcal R|.
\]
Eliminating $u$ gives
\begin{equation}\label{gauss:eq:real-corepenalty}
 e\le2h+2f-4+\tfrac32r+t-\sum_{v\in S}P_v,
 \qquad
 P_v=\frac{3s_v+d_v-4}{2}\ge2(s_v-1)\ge0.
\end{equation}
Here $\sum_v s_v=r$, and the last inequality uses $d_v\ge s_v$.

The block graph $B$ has $b=r+t$ vertices. Every nonempty block
retains a core occurrence. Suppression only replaces a collection
of edges joining the same two blocks by one such edge. Thus the
original and compressed block graphs have the same components and
the same bipartiteness. Color every core edge by
its connected component in $B$, and let $c_B$ be the number of
components. Around a vertex, colors can change only at a sign-change
corner: the two boundary occurrences meeting at any other corner
lie in the same block. A vertex touching $k_v\ge2$ colors therefore
has $s_v\ge k_v$ and costs $P_v\ge2(k_v-1)$. Vertices outside $S$
are monochromatic. The incidence graph of core vertices and colors
is connected, because the core graph is connected. Thus
$\sum_v(k_v-1)\ge c_B-1$, and
\begin{equation}\label{gauss:eq:real-colored}
 e\le2h+2f-4+\tfrac32r+t-2(c_B-1).
\end{equation}
This also holds for nonorientable maps, whose vertex links still
have a cyclic order up to reversal.

\emph{The rank of the length constraints.}
The block lengths are prescribed sums of the positive core-edge
lengths. Their matrix is the unsigned incidence matrix of $B$;
a loop has coefficient two. Let $c_b$ be the number of bipartite
components and $c_o=c_B-c_b$ the number of nonbipartite components.
Its left kernel consists of vectors $y$ satisfying $y_i+y_j=0$
on every nonloop edge and $2y_i=0$ at every loop. A connected
bipartite component contributes one kernel dimension; an odd cycle,
including a loop, forces the vector to vanish on its component.
The rank is therefore exactly $b-c_b$ over $\mathbb R$.
The same incidence system also reconstructs the lengths. In each
bipartite component choose a spanning tree by the order of first edge
occurrence. Given the other edge lengths, eliminate leaves. The final equation is the
signed balance between the two parts. In a nonbipartite component,
choose deterministically a spanning tree together with an edge
forming an odd cycle. Eliminate off-cycle leaves; the unsigned
odd-cycle system has determinant of absolute value two and has
at most one real solution. A loop is the one-edge instance. A
determinant of absolute value two may obstruct integrality, but
the real solution remains unique. Discard solutions that are
nonintegral or nonpositive. The deterministic choice of these
subgraphs introduces no counting factor.

There are $d=e-b+c_b\ge0$ free lengths. Their positive sum is at
most $n$, and the number of ordered assignments is at most
$\binom nd\le n^d/d!$, with one assignment when $d=0$.
Equation~\eqref{gauss:eq:real-colored} yields
\begin{equation}\label{gauss:eq:real-d}
 d\le D-(c_B-1)-c_o.
\end{equation}
In particular $d\le D$, and $d\le D-1$ whenever $B$ is not
bipartite. Also $e-d=b-c_b\le R+2$ and
$e\le2h+O_R(1)$. Cases with a negative upper bound for $d$ are empty.

\emph{Counting cores uniformly in $h$.}
For fixed $e$, there are at most $(2e)^{f-1}$ choices of positive
face lengths. The rooted lists are then labelled, and all pairings
and sign assignments are bounded by
\[
 (2e-1)!!\,2^{2e}=2^e\frac{(2e)!}{e!}.
\]
Signs already determine the twists. For $D'=D$ in the bipartite
case and $D'=D-1$ in the other case, the length count obeys
\[
 \frac{n^d}{d!}
 \le\frac{(e+1)^{R+2}}{e!}(n+1)^{D'}.
\]
The factorial from the free lengths compensates for the number of
cores, since
$(2e)!/(e!)^2\le4^e$. Summing the remaining exponential times a
fixed polynomial over $e\le2h+O_R(1)$ gives $K_R^{h+1}$.
The ordered rooted core lists already record the blocks in their
prescribed order. Their signs identify the sign-change corners,
and each first position identifies its root. Thus the vertex marks,
block labels and roots are all recovered from the same data. This
proves both counts. Coloring each component of a bipartite $B$ by
two signs and
adding its incidence equations with those signs proves the asserted
balance.
\end{proof}

The bipartition balance in the lemma can differ from the balance
between the original $+$ and $-$ signs. Indeed, real Gaussian
matrices already give a contribution of order $1/N$ to unbalanced
words such as $G_N^2$. The coefficient sum must therefore accommodate
all bipartition balances.

\begin{lemma}[Coefficient summation]\label{gauss:lem:real-poisson}
Let $A_1,\ldots,A_q$ be independent Poisson variables of means
$\mu_1,\ldots,\mu_q$, with $\sum_j\mu_j=2u$. For every fixed
$\eta\in\{1,-1\}^q$,
\[
 \Prob\left(\sum_j\eta_jA_j=0\right)
 \le e^{-2u}I_0(2u)\le C(1+u)^{-1/2}.
\]
Consequently the same bound, with constant $2^q C$, holds for the
union over all sign vectors.
\end{lemma}
\begin{proof}
The positive and negative sums are independent Poisson variables of
means $\mu_+$ and $\mu_-$, where $\mu_++\mu_-=2u$. Their equality
probability is $e^{-2u}I_0(2\sqrt{\mu_+\mu_-})$.
The power series for $I_0$ has nonnegative coefficients, and
$\sqrt{\mu_+\mu_-}\le u$. The angular integral representation of $I_0$
gives the last bound. The formula includes a zero mean on either
side. In an exponential expansion, the actual blocks may merge
after zero displayed blocks are deleted. Transfer the sign of an
actual block to every displayed block it contains, and extend it
arbitrarily to the zero blocks. Its balance then belongs to the
stated finite union. Removing the conditions that those blocks
vanish only enlarges that event.
\end{proof}

\begin{proposition}[Growing-time trace comparison]\label{gauss:prop:fnm-traces}
For every fixed integer $p\ge1$, there are constants $B_p,C_p',c_p>0$
such that the following holds. Let $\delta\ge0$, $T\ge1$,
$K=\lceil100(p+1)T\rceil$, and assume
\[
 B_p(K+1)^2/N\le1/2,\qquad T\le N/(64p).
\]
Then, with $\zeta_p=C_p'e^{-c_pT}+C_p'e^{-c_pN}$,
\begin{align}
 \left|\E\tau_N[(Q^G_{N,\delta,T})^p]-\tau(q_{\delta,T}^p)\right|
   &\le C_p'N^{-1}T^{2p+1/2}+\zeta_p,\label{gauss:eq:real-mean}\\
 \operatorname{Var}\!\left(\tau_N[(Q^G_{N,\delta,T})^p]\right)
   &\le C_p'N^{-2}T^{4p+1/2}+\zeta_p.\label{gauss:eq:real-var}
\end{align}
The constants are uniform in $\delta$.
\end{proposition}

\begin{proof}
We first truncate each exponential at degree $K$:
\[
 P_t=e^{-(1+\delta)t}\sum_{j=0}^K\frac{t^jG_N^j}{j!},\qquad
 A_N=\int_0^TP_tP_t^{\mathsf T}\dd t,
\]
and let $A_c$ be its circular counterpart. All scalar coefficients
and real Wick contributions are nonnegative. A word in $A_N^p$
has at most $2p$ sign changes and $pK$ pairs. The $h=0$ one-face
term equals its circular moment; the empty word contributes one
in both models. For $h\ge1$, Lemma
\ref{gauss:lem:real-count}, together with \eqref{gauss:eq:real-weight}, gives
a geometric series with ratio at most $B_p(K+1)^2/N$. Summing it
at this finite degree gives per-word bounds
\[
 C_p'N^{-1}(K+1)^{p+1}\quad\text{for bipartite cores},\qquad
 C_p'N^{-1}(K+1)^p\quad\text{for the remaining cores}.
\]
In the first class, the balances of different pairings need not
agree. Write $a$ for the vector of the $q=2p$ displayed lengths,
and $M_{\mathrm{bip}}(a)$ for their total weighted nonleading
bipartite contribution. The preceding bound implies the pointwise
inequality
\begin{equation}\label{gauss:eq:balance-union}
 M_{\mathrm{bip}}(a)
 \le C_p'N^{-1}(K+1)^{p+1}
       \sum_{\eta\in\{1,-1\}^{q}}\one_{\{\eta\cdot a=0\}}.
\end{equation}
Indeed if the contribution is nonzero, at least one displayed
balance holds by Lemma~\ref{gauss:lem:real-count} and the zero-block
extension in Lemma~\ref{gauss:lem:real-poisson}; otherwise the
inequality is immediate. The union over sign vectors accommodates
the different bipartitions with a factor depending only on $p$, including when
some displayed blocks vanish.

For $m$ time variables put $u=\sum_i t_i$. After multiplication by
$e^{-2u}$, the full scalar coefficient sum is the distribution of
$2m$ independent Poisson block lengths with means
$t_1,t_1,\ldots,t_m,t_m$. Extending the truncated nonnegative sum
can only increase it. Lemma~\ref{gauss:lem:real-poisson} gives
$(1+u)^{-1/2}$ for the bipartite contribution, and
\[
 \int_{[0,T]^m}(1+u)^{-1/2}\dd t\le C_mT^{m-1/2}.
\]
No balance is used in the nonbipartite contribution; its coefficient
sum is at most one. For $m=p$ these two contributions are respectively
$O_p(N^{-1}T^{2p+1/2})$ and $O_p(N^{-1}T^{2p})$.
The damping factor $e^{-2\delta u}$ only improves the bounds.

In the covariance of two normalized traces, disconnected Wick
pairings cancel exactly. Connected maps have $f=2$, at most $4p$
sign changes and $2pK$ pairs; their weights are $N^{-2-h}$.
Summing over $h\ge0$ gives the per-word bounds
$C_p'N^{-2}(K+1)^{2p+1}$ and $C_p'N^{-2}(K+1)^{2p}$
in the two respective classes. The analogue of \eqref{gauss:eq:balance-union} now has $q=4p$.
With $m=2p$, the preceding coefficient argument gives $C_p'N^{-2}T^{4p+1/2}$ for the variance.
A face of degree zero has deterministic trace and zero covariance;
pure nonempty faces are covered by the count. In both estimates the
deficit sum is taken at finite Taylor degree; only its nonnegative scalar majorant is extended to the
full Poisson sum.

It remains to pass from the polynomial approximation to the
exponentials. Two $1/8$-nets of the real unit sphere have size at most $17^N$ each.
The net estimate for the operator norm and a Gaussian tail give
\[
 \Prob(\norm{G_N}>x)
 \le2\,17^{2N}e^{-9Nx^2/32}\le e^{-Nx^2/8},\qquad x\ge10.
\]
On $\norm{G_N}\le10$, uniformly for $t\le T$,
\[
 \norm{e^{-(1+\delta)t}e^{tG_N}-P_t}
 \le e^{9T}\left(\frac{10eT}{K+1}\right)^{K+1}
 \le e^{-100(p+1)T}.
\]
Both factors have norm at most $e^{9T}$. Telescoping the $p$th
powers therefore bounds the trace approximation error by
$2pT^p e^{-(82p+109)T}$. The deterministic circular error has the
same bound, since $\norm c=2$.
On the complementary event, with $R_N=\norm{G_N}$, both Gramians
have norm at most $Te^{2TR_N}$ and the squared trace error is at
most $4T^{2p}e^{4pTR_N}$. Integrating this expression against the
norm tail, using $T\le N/(64p)$, bounds its expectation on
$\{R_N>10\}$ by $C_p'e^{-c_pN}$. Thus the trace approximation
error has both its $L^1$ norm and its squared $L^2$ norm bounded
by $\zeta_p$. The mean estimate transfers by $L^1$ control, and
$\operatorname{Var}Z\le2\operatorname{Var}W+2\E|Z-W|^2$
transfers the variance estimate.
\end{proof}

\section{Uniform counting with odd multiplicities}
\label{app:physical-reconstruction}

Nonzero odd moments contribute through index patterns in which some
ordered entries occur an odd number of times. We count these patterns
by dividing their occurrences into pairs and triples. The resulting
minimal endpoint quotient retains enough information to recover every
vertex after degree-two chains have been compressed. Its core count
can then be combined with the factorial from the prescribed block
lengths. A final forest count restores the physical vertex
identifications.

\subsection{Pair and triple refinements}

A repeated entry can be divided into pairs, with one triple left over
when its multiplicity is odd. This refinement encodes the index pattern;
its moment weight is still determined by the original physical entries.
The classes are distinguished by their occurrence sets. Two classes
with the same ordered endpoints remain distinct edges until the
physical vertex labels are restored.

Consider $f\in\{1,2\}$ prescribed nonempty rooted cyclic sign words,
with total length $\ell$. Each word is displayed as a concatenation
$\sigma_1^{k_1}\cdots\sigma_b^{k_b}$, where $\sigma_j\in\{+,-\}$
and $k_j\ge0$; its root is the corner immediately before the first
displayed block. The block labels remain distinct even when consecutive
blocks have the same sign. Order the positions first by face and then
by rooted position. Let $\pi$ be a partition of these positions into
classes of size two or three, numbered by their first occurrence.
Associate to each class $a$ two flags
$a_{\rm r},a_{\rm c}$. At a positive occurrence the boundary goes
from $a_{\rm r}$ to $a_{\rm c}$, and at a negative occurrence it
goes in the opposite direction. On each face, identify the ending flag
of every occurrence with the starting flag of the next occurrence,
including the last and first. The equivalence relation generated by
these identifications is the \emph{minimal endpoint quotient} of
$\pi$. Its classes form the vertices of a directed multigraph
$\Gamma_\pi$, with one edge for each class of $\pi$. Parallel edges,
including edges with equal ordered endpoints, keep their distinct
class identifiers.

We consider partitions for which $\Gamma_\pi$ is connected. Write
$V_\pi$ for its number of vertices and $n$ for its number of edges.
If $s$ classes have size three, then
\[
 n=\frac{\ell-s}{2},\qquad g=n-V_\pi+1.
\]
The closed words together traverse each edge twice or three times.
Their odd-edge subgraph has even degree at every vertex. Thus $s\ge1$
implies $g\ge1$, because a nonempty even-degree multigraph contains
a cycle.

\begin{lemma}[Minimal pair--triple cores]
\label{odd:lem:pair-triple-core}
Fix an integer $p\ge1$ and $f\in\{1,2\}$. Consider $f$ prescribed nonempty rooted
cyclic sign words, each with at most $2p$ displayed constant-sign blocks.
Partition their $\ell$ occurrences into classes of size two or three,
with at least one class of size three. Impose equality of the physical
row endpoints within every class, and equality of its physical column
endpoints. Let $H$ be the resulting minimal endpoint quotient: each
class is a separate directed edge, even when two such edges have the
same ordered endpoints. Assume that $H$ is connected. If $n$ is its
number of edges, $V$ its number of vertices, and
$g=n-V+1$, then $g\ge1$.

Remove empty displayed blocks and retain every boundary between the
remaining blocks, including boundaries between blocks of the same sign.
Keep the original roots. Mark the quotient vertices containing a block
boundary; this includes every root, also for pure words. Suppress all unmarked vertices of degree
two, counting a loop twice in the degree. The resulting connected core
has $E$ chains and $M$ occurrence positions in its $f$ rooted face lists.
Every chain has multiplicity two or three, and each of its occurrences
is contained in one face and one block. If $q$ chains have multiplicity
three, then
\[
 M=2E+q,\qquad E\le2g+3pf-2.
\]
The endpoint quotient of the compressed occurrence partition is still
minimal. Thus the rooted face lists, their signs and block labels, and
the partition of their positions into pair and triple chain classes
already determine every core vertex; no further endpoint partition is
needed.

Let $B_{ja}$ be the number of occurrences of chain $a$ in original
block $j$, and put $\rho=\operatorname{rank}_{\mathbb R}B$ and
$F=E-\rho$. Then
\begin{equation}\label{odd:eq:pair-triple-free-lengths}
 0\le\rho\le2pf,\qquad F\le2g+pf+f-2.
\end{equation}
For a fixed core, the original block lengths $k$ impose $Bx=k$ on its
positive chain lengths. There are at most
\begin{equation}\label{odd:eq:pair-triple-length-count}
 \binom{\lfloor\ell/2\rfloor}{F}
 \le\frac{(\ell/2)^F}{F!}
\end{equation}
possible length vectors. Each such vector has at most one expansion to
the original pair--triple occurrence partition. If the original
partition has $s$ triple classes, then $\sum_{a:\,m_a=3}x_a=s$ and,
in particular, $1\le q\le s$.
\end{lemma}
\begin{proof}
For each occurrence class $a$, introduce two distinct nodes $(a,R)$
and $(a,C)$ for its row and column endpoints. A positive occurrence
starts at $(a,R)$ and ends at $(a,C)$; a negative occurrence starts at
$(a,C)$ and ends at $(a,R)$. Every boundary corner gives an undirected
edge joining the end node of its preceding occurrence to the start
node of its following occurrence. Denote this auxiliary graph by
$\mathcal L$. Its connected components are exactly the formal vertices
of the minimal endpoint quotient. At each vertex of $H$, its component
of $\mathcal L$ will be called its link. The two nodes of a class are
distinct even when they lie in the same link and its edge is a loop.
Each endpoint node has degree two or three, according to its class
size.

The quotient of each original cyclic walk is connected. By assumption
their union, and hence $H$, is connected. Its weighted occurrence
degree is even at every vertex.
Reducing those degrees modulo two shows that its triple edges form a
nonempty even-degree subgraph. This subgraph contains a cycle, including
the possibility of a loop or a pair of parallel edges. Hence $g\ge1$.

At an unmarked vertex all its corners lie inside constant-sign blocks.
Each edge of its link consequently joins a row node to a column node.
The link is connected by minimality, and is bipartite by these two types.
There can be no unmarked vertex of degree one in $H$: its link would have just
one node and could not have its required positive degree. If its degree
is two, its link has two nodes of opposite types joined by either two
or three parallel corner edges. Their class multiplicities are equal,
and the corners give a bijection between their occurrences, preserving
the face, the block label and the sign. These nodes cannot be the two
ends of a single edge: that would make this loop the whole connected
graph, and its vertex would contain a marked root. Thus suppression
joins two distinct edges in a directed chain.

An unmarked vertex of degree three is also impossible. Its connected
bipartite link would have one node in one part and two in the other;
the degree of the single node would have to equal the sum of the other
two degrees. The former is at most three and the latter at least four.
Every remaining unmarked vertex therefore has degree at least four.
There is no component consisting of an entirely unmarked cycle, because
it would be the whole connected graph and contain a face root. Maximal
chains consequently end at retained vertices, possibly at the same
vertex twice. The occurrence bijections along a chain give its two or
three contiguous strands. No strand crosses a retained corner, so it
belongs to one original face and block. Suppression preserves the number
of faces and the cycle rank.

Compression also preserves the minimal endpoint quotient. Suppressing
a degree-two
vertex removes its entire two-node link. At a retained vertex, the
endpoint nodes of the incident unsuppressed edges are simply renamed
as the endpoint nodes of their chains. This renaming is one-to-one,
including when a chain has equal endpoints: its row and column nodes
remain distinct. No corner at a retained vertex is removed. Thus its
link graph is unchanged up to renaming. It therefore remains
connected when the endpoint equalities are computed from the
compressed occurrence partition. The chain classes and boundary
corners consequently generate exactly the retained vertices.

Conversely, expand a core chain of length $x$ at all its occurrences.
For a positive occurrence use its $x$ directed positions in increasing
order, and for a negative occurrence use them in decreasing order.
For each $1\le j\le x$, group the $j$th directed positions across all
two or three strands into one class. These are the unique classes
that produce the directed subdivision. Their interior vertices have
the two-node links just described. The original retained links are
unchanged. The first positions in the rooted face lists fix all roots.
This reconstruction applies to loops, parallel chains, and chains occurring
in both faces. The directed positions determine every matching at an
interior vertex, while the retained face roots and block labels fix
the positions in the original words.

Let $m$ be the number of marked vertices, $u$ the number of other core
vertices, and $D_M$ the sum of their marked degrees. If $b$ is the total
number of blocks after the stated preprocessing, then $m\le b\le2pf$.
The degree bound and Euler identity give
\[
 2E\ge4u+D_M,\qquad u=E-g-m+1,
\]
and hence
\[
 E\le2g+2m-2-D_M/2\le2g+\tfrac32m-2
       \le2g+3pf-2.
\]
Here $D_M\ge m$, since every marked vertex has positive degree.

Give each chain its physical row-to-column direction, and let $\partial$
be the oriented vertex-edge incidence matrix of the core. If $\sigma_j$
is the sign of block $j$, then
\[
 \partial B^{\mathsf T}\operatorname{diag}(\sigma_j)=J,
\]
where column $j$ of $J$ is the endpoint difference of that block.
The graph with those blocks as edges and the marked vertices as
vertices is a union of $f$ closed walks. Each of its connected
components contains at least one of those walks, so it has at most
$f$ components. Consequently
$\rho\ge\operatorname{rank}J\ge m-f$. Together with $\rho\le b$ this
implies
\[
 F\le E-m+f
   \le2g+m+f-2-D_M/2
   \le2g+pf+f-2.
\]

Choose a maximal linearly independent set of columns of $B$ by a fixed
rule. The remaining $F$ lengths determine the pivot lengths uniquely
over the reals. Discard nonpositive or nonintegral solutions. Since
each chain has at least two occurrences, every admissible vector
satisfies $\sum_a x_a\le\ell/2$. The positive free lengths therefore
have at most $\binom{\lfloor\ell/2\rfloor}{F}$ assignments. The unique
expansion already established completes the counting assertion. Each
triple chain of length $x_a$ expands into exactly $x_a$ triple classes,
proving the final identity and the bounds on $q$.
\end{proof}

\begin{lemma}[Counting pair--triple partitions]
\label{odd:lem:pair-triple-count}
Fix an integer $p\ge1$ and $f\in\{1,2\}$. For $f$ prescribed nonempty
rooted cyclic sign words, each with at most $2p$ displayed constant-sign
blocks, the number of pair--triple partitions with exactly $s\ge1$
triples and connected minimal endpoint quotient of cycle rank $g\ge1$
is at most
\begin{equation}\label{odd:eq:pair-triple-count}
 C_p^g(1+\ell)^{pf+2g+s+f-3}.
\end{equation}
The constant is independent of the words, their displayed block lengths,
and the integers $g,s$.
\end{lemma}
\begin{proof}
Use the core of Lemma~\ref{odd:lem:pair-triple-core}. For fixed
$E,q$, its $M=2E+q$ occurrence positions are numbered in rooted face
order, and its chain classes are numbered by their first occurrence.
Dividing these positions into $f$ nonempty consecutive face lists
costs at most $M^{f-1}$. The number of partitions of the positions into
$E-q$ pairs and $q$ triples is
\[
 \frac{M!}{2^{E-q}6^q(E-q)!q!}\le8^EE!M^q.
\]
Indeed, $M!\le(2E)!M^q$, and the remaining factor satisfies
\[
 \frac{(2E)!}{2^EE!}\frac{\binom Eq}{3^q}
 \le2^EE!\left(\frac43\right)^E\le8^EE!.
\]
Assigning a sign and one of at most $2pf$ block labels to each occurrence
costs at most $(4pf)^M$. These data determine the minimal endpoint
quotient. Discard records with the wrong block order, disconnected
quotient, wrong cycle rank, or an unsuppressed unmarked degree-two
vertex. Since $M\le3E$ and $f\le2$, the number of cores is at most
$C_p^EM^{f-1}E!M^q$.

The free-length dimension $F$ may vary among these cores. For each
individual core, Lemma~\ref{odd:lem:pair-triple-core} gives
\[
 \frac{E!}{F!}=\prod_{j=0}^{\rho-1}(E-j)
       \le E^\rho\le E^{2pf},\qquad F\le pf+2g+f-2.
\]
In particular, its number of admissible length vectors is at most
\[
 \frac{E^{2pf}}{E!}(1+\ell)^{pf+2g+f-2}.
\]
The resulting bound is uniform over the possible ranks.
Multiplying by the core count bounds the contribution for fixed $E,q$ by
\[
 C_p^EM^{f-1}E^{2pf}M^q
 (1+\ell)^{pf+2g+f-2}.
\]
Only cores with $1\le q\le s$ and $M\le\ell$ can contribute.
For these cores,
\[
 M^q\le M(1+\ell)^{s-1}\le3E(1+\ell)^{s-1}.
\]
Sum over $1\le E\le2g+3pf-2$ and $1\le q\le E$. After removing
the displayed power of $1+\ell$, the sum is bounded by
\[
 \sum_{E=1}^{2g+3pf-2} C_p^E E^{2pf+f+1}\le C_p'{}^g.
\]
Indeed, $E\le(2+3pf)g$ for $g\ge1$, and every fixed power of $g$
is bounded by an exponential in $g$. Both the polynomial degree and
the summation range depend only on $p$ and $g$, so the resulting constant is uniform in $s$.
This proves \eqref{odd:eq:pair-triple-count}.

\end{proof}

\begin{theorem}[A uniform count for odd entry patterns]
\label{odd:thm:uniform-odd-count}
Fix an integer $p\ge1$ and $f\in\{1,2\}$. Consider $f$ prescribed
nonempty rooted cyclic sign words, each with at most $2p$ displayed
constant-sign blocks, with total length $\ell$. Count canonical
physical-index patterns in which every ordered entry occurs at least
twice and exactly $s\ge1$ entries have odd multiplicity. When $f=2$,
require that the two words share at least one physical ordered entry.
If $v$ is the number of physical vertices and
\[
 d=\frac{\ell-s}{2}+1-v,
\]
then their number is at most
\begin{equation}\label{odd:eq:uniform-odd-count}
 C_p^d(1+\ell)^{pf+2d+s+f-3}.
\end{equation}
The constant is uniform in $d,s,\ell$ and all displayed block lengths,
including zero lengths.
\end{theorem}
\begin{proof}
Order the physical entry classes and the occurrences within each class
by their first appearance in the rooted face lists. For one face, take
the first three occurrences of each odd class as a triple, then pair
the remaining occurrences consecutively; pair each even class
consecutively as well.
For two faces, first choose the earliest class meeting both faces.
If its total multiplicity is even, use its first occurrence in each
face as a cross-face pair. If the total multiplicity is odd, add the
earliest remaining occurrence to form a cross-face triple. Pair the
remaining occurrences of this class consecutively, and refine all
other classes by the one-face rule in the combined occurrence order.
This produces a deterministic partition $\pi$ with
exactly $s$ triples and $n=(\ell-s)/2$ abstract edges. Its minimal
endpoint quotient is connected: each individual face traces a connected
walk, and the selected cross-face class joins the two walks.

The actual physical vertex labels satisfy all endpoint relations
generating $\Gamma_\pi$. They therefore partition its $V_\pi$ formal
vertices into $v$ physical vertices. Write $c=V_\pi-v$. Since the
minimal graph has cycle rank $g=n-V_\pi+1\ge1$,
\begin{equation}\label{odd:eq:refined-physical-deficit}
 g+c=d.
\end{equation}
The pair consisting of $\pi$ and this vertex partition uniquely recovers
the original physical index-equality pattern: identify the formal vertex
of each word corner, apply the partition, and label the resulting classes
in order of first visit. Its physical ordered-entry classes are then
exactly the classes with equal ordered endpoint pairs. The deterministic
refinement is therefore an injective encoding of the physical index
pattern.

For a fixed $\pi$, label its formal vertices in order of first visit.
There are at most
\[
 \binom{\binom{V_\pi}{2}}{c}
 \le\frac{V_\pi^{2c}}{c!}
\]
partitions with $c$ identifications. Within each class, joining its least
labelled vertex to every other vertex gives a canonical forest with $c$
edges; its connected components recover that partition. Some coarsenings
merge several triple classes into one physical entry or change its
parity. Including these records only increases the bound.

Since $V_\pi\le n\le\ell$, Lemma~\ref{odd:lem:pair-triple-count}
and \eqref{odd:eq:refined-physical-deficit} bound the required count by
\[
 (1+\ell)^{pf+2d+s+f-3}
 \sum_{g=1}^{d}\frac{C_p^g}{(d-g)!}
 \le C_p^d(1+\ell)^{pf+2d+s+f-3},
\]
after enlarging $C_p$. This also proves $d\ge1$. Empty displayed
blocks can be removed without increasing the block count. Pure words
are included by retaining their roots throughout compression.
\end{proof}

\section{The upper spectral edge of the circular covariance}
\label{sec:free-edge}

The limiting measure in Theorem~\ref{free:thm:main} has right endpoint
$1/2$. To identify the norm of the rescaled covariance, we also need
an upper moment bound whose order can grow as the damping vanishes.
The estimate below preserves the exponential rate in the moment order;
its polynomial loss in the damping parameter disappears after taking
the moment root.

\begin{theorem}\label{edge:thm:free-norm}
For a standard circular element in a faithful tracial von Neumann
algebra,
\begin{equation}\label{edge:eq:norm-limit}
 \lim_{\delta\downarrow0}\delta^2\norm{q_\delta}=\frac12.
\end{equation}
\end{theorem}

A common random translation bounds all the block-length fluctuations
at once, giving a moment estimate with an absolute constant. We then
approximate the covariance by a polynomial and apply a free Wick
polynomial inequality to turn the moment estimate into a norm bound.

\subsection{A bound uniform in the moment order}

\begin{lemma}\label{edge:lem:translation}
For every integer $p\ge1$ and $\ell\in\N_0^p$,
\begin{equation}\label{edge:eq:majorization}
 \Phi_p(\ell)\le F_p(\ell+\mathbf1).
\end{equation}
The function $F_p$ is increasing in each coordinate. There are
nonnegative functions $G_{p,k}$, homogeneous of degree $p-1-k$, such that
\begin{equation}\label{edge:eq:translation}
 F_p(t+s\mathbf1)=\sum_{k=0}^{p-1}G_{p,k}(t)s^k,
 \qquad t\in\R_+^p,\quad s\ge0.
\end{equation}
Writing $T=\sum_i t_i$ and
\[
 a_{p,k}=\int_{\R_+^p}e^{-T}G_{p,k}(t)\dd t,
\]
we have $a_{1,0}=1$ and, for $p\ge2$,
\begin{equation}\label{edge:eq:coefficients}
 a_{p,k}=\frac{p^{k-1}}{k!}
      \binom{2p-k-2}{p-1-k}2^{-(p-1-k)},
 \qquad 0\le k\le p-1.
\end{equation}
In particular,
\begin{equation}\label{edge:eq:coefficient-ratio}
 a_{p,0}=\frac{\Cat_{p-1}}{2^{p-1}},\qquad
 \frac{a_{p,k}}{a_{p,0}}\le\frac{p^k}{k!}.
\end{equation}
\end{lemma}

\begin{proof}
For $p=1$, \eqref{edge:eq:majorization} is an equality. For larger $p$,
rotate a minimum of $\ell$ to the first coordinate in
\eqref{free:eq:exact-rec}. Apply the induction hypothesis to each
residual interval and use
\[
 \binom{\ell_1+1}{k}\le\frac{(\ell_1+1)^k}{k!}.
\]
The resulting sum is precisely the recursion for
$F_p(\ell+\mathbf1)$. A common translation preserves the choice of a
minimum at every recursive call, including when minima are tied.

To increase one coordinate of a locally symmetric word by an integer
$r\ge0$, insert $+^r-^r$ at the corresponding peak and match the new
letters in nested order. This gives an injection of pairing sets.
Thus $\Phi_p$ is increasing in each coordinate. Integer dilation and
Lemma~\ref{free:lem:F} give the same conclusion for $F_p$ at rational
points, and continuity extends it to all real points.

For fixed $t$, apply \eqref{free:eq:F-rec} to $t+s\mathbf1$.
Each factor at a minimum is a power of $t_i+s$ with a positive
coefficient. Induction in the residual intervals proves
\eqref{edge:eq:translation} with nonnegative coefficients. Homogeneity
of $F_p$ gives the degree of $G_{p,k}$. On an ordering wall the
polynomial expressions agree for every $s\ge0$, by continuity of
$F_p$. They therefore agree coefficient by coefficient, so the
functions $G_{p,k}$ are well defined there.

Integrating \eqref{edge:eq:translation} against $e^{-T}$ identifies
$a_{p,k}$ with the coefficient of $s^k$ in $A_p(s)$ from
\eqref{free:eq:lagrange}. Expansion of that formula proves
\eqref{edge:eq:coefficients}. For $k\ge1$, the ratio to its constant
coefficient is
\[
 \frac{a_{p,k}}{a_{p,0}}
 =\frac{p^k}{k!}
       \prod_{j=0}^{k-1}\frac{2(p-1-j)}{2p-2-j}
 \le\frac{p^k}{k!}.
\]
The formula for $a_{p,0}$ is \eqref{free:eq:J}.
\end{proof}

\begin{lemma}\label{edge:lem:poisson-maximum}
Let $N$ be a rate-one Poisson process and put
$D_T=\sup_{0\le s\le T}\abs{N(s)-s}$. There is an absolute constant
$C$ such that, for every integer $k\ge1$ and $T\ge0$,
\begin{equation}\label{edge:eq:maximal-moment}
 \E D_T^k\le C^k\bigl((kT)^{k/2}+k^k\bigr).
\end{equation}
\end{lemma}

\begin{proof}
For real $\theta$, the process
\[
 \exp\{\theta(N(s)-s)-s(e^\theta-1-\theta)\}
\]
is a positive martingale of mean one. For $0<\theta\le1$, both
$e^\theta-1-\theta$ and $e^{-\theta}-1+\theta$ are at most $\theta^2$.
Stopping at the first upward or downward crossing of level $x$ and
using the maximal inequality for a positive martingale gives
\[
 \Prob(D_T\ge x)\le 2\exp\{-\theta x+T\theta^2\}.
\]
For $T>0$, choose $\theta=\min\{x/(2T),1/2\}$. Hence, for an absolute
$c>0$,
\[
 \Prob(D_T\ge x)
 \le 2\exp\{-c\min(x^2/T,x)\}
 \le 2e^{-cx^2/T}+2e^{-cx}.
\]
Integrating $k x^{k-1}$ against this bound gives gamma integrals of
orders $k/2$ and $k$. Their elementary factorial bounds prove
\eqref{edge:eq:maximal-moment}. For $T=0$, the left side is zero.
\end{proof}

\begin{proposition}\label{edge:prop:all-moments}
There is an absolute constant $C$ such that, for every integer $p\ge1$
and every $0<\delta\le1$,
\begin{equation}\label{edge:eq:catalan-bound}
 \tau(q_\delta^p)
 \le \frac{\Cat_{p-1}}{2^{p-1}(2\delta)^{2p-1}}
       \left\{e^{Cp^2\sqrt\delta}+e^{Cp^2\delta}-1\right\}.
\end{equation}
Consequently, after enlarging $C$ if necessary,
\begin{equation}\label{edge:eq:all-moments}
 \delta^{2p}\tau(q_\delta^p)
       \le 2\delta\,2^{-p}e^{Cp^2\sqrt\delta}.
\end{equation}
\end{proposition}

\begin{proof}
Fix a time vector $t\in\R_+^p$, put $T_i=\sum_{j\le i}t_j$, and let
$N_1,N_2$ be independent rate-one Poisson processes. Their increments
on $[T_{i-1},T_i]$ give the independent block lengths $A_i,B_i$ in
\eqref{free:eq:poisson}. Write
\[
 D_j=\sup_{0\le s\le T}\abs{N_j(s)-s},\qquad
 Z=1+4D_1+2D_2,\qquad T=T_p.
\]
Writing $E_j(s)=N_j(s)-s$, the block and valley identities are
\[
 A_i=t_i+E_1(T_i)-E_1(T_{i-1}),\qquad
 v_i=E_1(T_i)-E_2(T_i).
\]
On the event of total balance, these give
$\abs{v_i}\le D_1+D_2$ and $A_i\le t_i+2D_1$. Therefore
\[
 h_i-\min_jv_j+1\le t_i+Z.
\]
Valley insertion, \eqref{edge:eq:majorization}, and coordinatewise
monotonicity give
\begin{equation}\label{edge:eq:random-shift}
 P_p(A,B)\le F_p(t+Z\mathbf1).
\end{equation}
Off the balance event the left side is zero, so this inequality holds
without conditioning. Lemma~\ref{edge:lem:poisson-maximum} and the
triangle inequality in $L^k$ imply
\begin{equation}\label{edge:eq:shift-moment}
 \E Z^k\le C^k\bigl((kT)^{k/2}+k^k\bigr),\qquad k\ge1.
\end{equation}
The constant is independent of the number of blocks: the two process
maxima control every block and every valley at once.

Insert \eqref{edge:eq:translation} into \eqref{edge:eq:random-shift}
and use the exact Poisson representation. For fixed $t$, the
coefficients $G_{p,k}(t)$ are deterministic, so
\[
 \E F_p(t+Z\mathbf1)
   =\sum_{k=0}^{p-1}G_{p,k}(t)\E Z^k.
\]
All terms are nonnegative.
To integrate them, homogeneity gives, for $r\ge0$,
\begin{equation}\label{edge:eq:radial}
 \int_{\R_+^p}e^{-2\delta T}T^rG_{p,k}(t)\dd t
 =a_{p,k}
   \frac{\Gamma(2p-1-k+r)}{\Gamma(2p-1-k)}
   (2\delta)^{-(2p-1-k+r)}.
\end{equation}
The term $k=0$ is exactly
\[
 M_p(\delta):=a_{p,0}(2\delta)^{-(2p-1)}.
\]
For $1\le k\le p-1$, use \eqref{edge:eq:shift-moment} and
\eqref{edge:eq:radial} with $r=k/2$ and $r=0$. The gamma ratio in the
first case is at most $(2p)^{k/2}$. For example, writing the logarithm
of this ratio as the integral of the digamma function and using
$\psi(x)\le\log x$ gives this bound, since its arguments are positive
and its upper endpoint is at most $2p$.
Together with \eqref{edge:eq:coefficient-ratio}, the contribution
divided by $M_p(\delta)$ is at most
\[
 \frac{(Cp^{3/2}\sqrt\delta)^k k^{k/2}}{k!}
 +\frac{(Cp\delta)^k k^k}{k!}
 \le
 \frac{(Cp^2\sqrt\delta)^k}{k!}
 +\frac{(Cp^2\delta)^k}{k!}.
\]
Sum over $k$, extend the two nonnegative sums to all positive integers,
and add the term $k=0$. This proves
\eqref{edge:eq:catalan-bound}. Finally,
$\Cat_{p-1}\le4^{p-1}$ and $\delta\le\sqrt\delta$ yield
\eqref{edge:eq:all-moments}.
\end{proof}

The base $1/2$ in this estimate comes from the exact translated
coefficients. Dropping the balance constraint in
\eqref{edge:eq:random-shift} changes the fixed-order prefactor from
$\delta^{3/2}$ to $\delta$, but the resulting loss in a moment root is
$\exp\{O(\abs{\log\delta}/p)\}$. An order growing faster than
$\abs{\log\delta}$ therefore makes this prefactor loss vanish.

\subsection{From moments to the operator norm}

The semicircular Haagerup inequality \cite{Bozejko}, in the
nonhomogeneous form of \cite[Theorem~1.1, Eq.~(1.2)]{Parraud}, relates
the norm of a polynomial to its second moment. Its dependence on the
degree is polynomial, so it can be combined with the preceding bound
at a moment order of size $(\log(1/\delta))^2$. We include the free
Wick proof.

\begin{lemma}\label{edge:lem:wick-norm}
If $P$ is a polynomial of total degree at most $D$ in a standard free
semicircular family, then
\begin{equation}\label{edge:eq:wick-norm}
 \norm P\le(D+1)^{3/2}\tau(P^*P)^{1/2}.
\end{equation}
The same inequality holds for a polynomial of degree at most $D$ in
$c,c^*$.
\end{lemma}

\begin{proof}
Use the full Fock space representation of the semicircular family.
Write the vacuum vector of $P$ as
\[
 P\Omega=\sum_{n=0}^D\xi_n,\qquad \xi_n\in\mathcal H^{\otimes n}.
\]
The free Wick polynomial $W(\xi_n)$ has vacuum vector $\xi_n$ and its
normal-order expansion has $n+1$ terms, indexed by the number of
creation operators. On tensor degrees at least the number of
annihilations, each term is a tensor flattening of $\xi_n$, with the
annihilation indices reversed, tensored with the identity on the
remaining coordinates; it vanishes on lower degrees. The norm of
the flattening is bounded by its Hilbert--Schmidt norm
$\norm{\xi_n}$. Thus
\[
 \norm{W(\xi_n)}\le(n+1)\norm{\xi_n}.
\]
The Wick decomposition gives $P=\sum_{n=0}^D W(\xi_n)$. Its orthogonal
vacuum components satisfy
$\sum_n\norm{\xi_n}^2=\tau(P^*P)$. Cauchy--Schwarz now gives
\[
 \norm P
 \le\left(\sum_{n=0}^D(n+1)^2\right)^{1/2}
       \tau(P^*P)^{1/2}
 \le(D+1)^{3/2}\tau(P^*P)^{1/2}.
\]
The norm of a polynomial is determined by its joint moments in a
faithful tracial representation, through
$\norm P=\lim_{m\to\infty}\tau((P^*P)^m)^{1/(2m)}$.
Thus the bound transfers from the Fock representation to the given
semicircular family.
Finally, $c=(s_1+is_2)/\sqrt2$ for two standard free semicircular
elements, and this substitution preserves total degree.
\end{proof}

\begin{proof}[Proof of Theorem~\ref{edge:thm:free-norm}]
For an integer $K\ge0$, put
\[
 S_K(t c)=\sum_{n=0}^K\frac{t^n c^n}{n!},\qquad
 q_{\delta,K}=\int_0^\infty e^{-2(1+\delta)t}
                         S_K(tc)S_K(tc)^*\dd t.
\]
This is a positive polynomial of degree at most $2K$. Its coefficient
expansion is obtained by retaining the terms $0\le n,m\le K$ in
\[
 q_\delta=\sum_{n,m\ge0}
 \frac{(n+m)!}{n!m![2(1+\delta)]^{n+m+1}}c^n(c^*)^m.
\]
The bound $\norm{c^n}\le\sqrt{e(n+1)}$ gives a summable norm
majorant, both for this series and for its termwise time integral.
All scalar coefficients and circular Wick traces in this expansion
are nonnegative. Thus, for every integer
$r\ge1$,
\begin{equation}\label{edge:eq:trunc-moments}
 \tau(q_{\delta,K}^r)\le\tau(q_\delta^r).
\end{equation}
Indeed, the truncation retains a subset of the terms in the absolutely
convergent trace expansion, for each $\delta>0$, $K$ and $r$.

Write $a=1+\delta$. The omitted terms have $n+m>K$. The bound
\[
 \sqrt{(n+1)(m+1)}\le\frac{n+m+2}{2}
\]
and the binomial sum at fixed $n+m$ give
\begin{equation}\label{edge:eq:trunc-norm}
 \norm{q_\delta-q_{\delta,K}}
 \le\frac e{4a}\sum_{j>K}(j+2)a^{-j}.
\end{equation}
The geometric sum can be evaluated explicitly:
\begin{equation}\label{edge:eq:trunc-scaled}
 \delta^2\norm{q_\delta-q_{\delta,K}}
 \le\frac e4(1+\delta)^{-K-1}\bigl[\delta(K+3)+1\bigr].
\end{equation}
Take $K=\lceil\delta^{-2}\rceil$. Since
$\log(1+\delta)\ge\delta/2$ for $0<\delta\le1$, this error is
$O(\delta^{-1}e^{-1/(2\delta)})$ and tends to zero.

For sufficiently small $\delta$, choose
$p=\lceil(\log(1/\delta))^2\rceil$. Apply
Lemma~\ref{edge:lem:wick-norm} to $q_{\delta,K}^p$ and use
\eqref{edge:eq:trunc-moments}. Positivity gives
\[
 \norm{q_{\delta,K}}
 \le(2Kp+1)^{3/(2p)}\tau(q_\delta^{2p})^{1/(2p)}.
\]
Proposition~\ref{edge:prop:all-moments} therefore yields
\[
 \delta^2\norm{q_{\delta,K}}
 \le\frac12(2Kp+1)^{3/(2p)}(2\delta)^{1/(2p)}
                         e^{C p\sqrt\delta}.
\]
Here $\log(2Kp+1)/p\to0$, $\abs{\log\delta}/p\to0$, and
$p\sqrt\delta\to0$. The truncation estimate proves
\[
 \limsup_{\delta\downarrow0}\delta^2\norm{q_\delta}\le\frac12.
\]

For the lower bound, fix an integer $r\ge1$. Positivity gives
\[
 \norm{q_\delta}\ge
             \frac{\tau(q_\delta^{r+1})}{\tau(q_\delta^r)}.
\]
The fixed-order asymptotics in Theorem~\ref{free:thm:main} imply
\[
 \liminf_{\delta\downarrow0}\delta^2\norm{q_\delta}
 \ge\frac{C_{r+1}}{C_r}
 =\frac{(2r-\tfrac32)(2r-\tfrac12)}{8r(r+1)}.
\]
Let $r\to\infty$ after this limit. The right side tends to $1/2$,
which proves \eqref{edge:eq:norm-limit}.
\end{proof}

\begin{corollary}\label{edge:cor:arbitrary-orders}
Put $b_\delta=\delta^2\norm{q_\delta}$. For every integer $p\ge1$,
every $\delta>0$, and every $0<T\le\infty$,
\begin{equation}\label{edge:eq:arbitrary-orders}
 \tau\bigl[(\delta^2q_{\delta,T})^p\bigr]
 \le \frac{\delta^2}{2\sqrt{(1+\delta)^2-1}}\,b_\delta^{p-1}.
\end{equation}
The inequality holds simultaneously for all $p$ and $T$, with the
same $b_\delta$. Its prefactor is asymptotic to
$\delta^{3/2}/(2\sqrt2)$, and $b_\delta\to1/2$ as $\delta\downarrow0$.
\end{corollary}
\begin{proof}
Positivity gives $0\le q_{\delta,T}\le q_\delta$, and hence
\[
 \tau(q_{\delta,T}^p)
 \le\norm{q_\delta}^{p-1}\tau(q_{\delta,T})
 \le\norm{q_\delta}^{p-1}\tau(q_\delta).
\]
Use the exact first moment from Section~\ref{sec:free} and
Theorem~\ref{edge:thm:free-norm}.
\end{proof}

\end{appendix}

\end{document}